\documentclass[
reprint,
superscriptaddress,
amsmath,
amssymb,
aps,
prd,
]{revtex4-2}

\usepackage{amsthm}
\theoremstyle{definition}

\usepackage[utf8]{inputenc}
\usepackage{amsmath}
\usepackage{slashed}
\usepackage{graphicx}
\usepackage{subcaption}    
\usepackage{xcolor}
\usepackage[pdftex]{hyperref}
\begin{document}
\title{Stringent requirements for detecting light-induced gravitational effects using interferometry}

\author{François Fillion-Gourdeau}
\affiliation{Infinite Potential Laboratories, Waterloo, Ontario, Canada, N2L 0A9}
\affiliation{Advanced Laser Light Source (ALLS) at INRS-EMT, 1650 blvd. Lionel-Boulet, Varennes, QC, J3X 1P7, Canada}

\author{Steve MacLean}
\affiliation{Infinite Potential Laboratories, Waterloo, Ontario, Canada, N2L 0A9}
\affiliation{Advanced Laser Light Source (ALLS) at INRS-EMT, 1650 blvd. Lionel-Boulet, Varennes, QC, J3X 1P7, Canada}

\begin{abstract}
	Intense laser fields have been proposed as a means to generate light-induced gravitational effects, providing a novel approach to investigate gravity and its coupling to electromagnetism in a controlled laboratory setting. In this article, a detection scheme based on interferometry is introduced to assess the feasibility of observing such effects. Initially, the space-time deformation and the resulting induced phase difference are evaluated in homogeneous electric fields. Using the theoretical minimum phase sensitivity bound --- a known result in quantum information --- and accounting for background signal coming from photon-photon scattering --- a fundamental quantum electrodynamics effect related to vacuum properties --- a set of stringent requirements for detectability is obtained. Then, a more realistic scenario is considered where gravitational effects are generated by an e-dipole pulse. In all cases considered, it is demonstrated that observing these effects presents significant challenges, even with the capabilities of current and foreseen laser infrastructures. 
\end{abstract}

\maketitle

\section{Introduction \label{sec:intro}}
According to general relativity, the curvature of spacetime is determined by the distribution of mass, momentum, and energy \cite{Misner:1973prb,Weinberg:1972kfs,Carroll:2004st}. This intricate relationship is captured in Einstein's equation, which directly links the Ricci tensor---a mathematical quantity that characterizes the geometry of spacetime---with the energy-momentum tensor, which contains all the information about the embedded matter and energy. In other words, general relativity describes a two-way interaction where spacetime influences how matter and energy behave, while matter and energy shape the geometry of spacetime. This nonlinear relationship is well-established both theoretically and experimentally, with numerous precise tests, including the perihelion precession of Mercury's orbit \cite{https://doi.org/10.1002/andp.19163540702,10.1119/1.1949625,Park_2017}, the gravitational bending of light \cite{PhysRevLett.92.121101,Fomalont_2009}, the gravitational redshift of light \cite{EARMAN1980175, PhysRevLett.45.2081,muller2010precision}, the direct observation of black holes \cite{Akiyama_2019}, and the direct detection of gravitational waves \cite{maggiore_book,PhysRevLett.116.061102,PhysRevLett.116.241103}.

Most of these tests involve astrophysical phenomena with celestial bodies having large masses, thus inducing strong gravitational fields. Incidentally, these tests probe the interaction between gravity and matter, but they are not as sensitive to other particles like photons. There is experimental evidence that demonstrates the effect of gravity on light, such as gravitational light bending and redshift. However, light-induced gravitational effects (LIGE) where strong electromagnetic fields curve spacetime and act as sources of gravitation, have not been observed. Nevertheless, according to Einstein's equation, such a phenomenon should exist because electromagnetic fields also carry energy and are therefore characterized by their own energy-momentum tensor. LIGE would be responsible for \textit{geons}, an unstable state of light held together by gravity \cite{PhysRev.97.511,PhysRev.105.1665}, and \textit{kugelblitze}, black holes created by the gravitational collapse of electromagnetic energy \cite{Senovilla_2015,PhysRevLett.133.041401}. LIGE is also very similar to the \textit{Gertsenshtein effect}, where electromagnetic waves are converted to gravitational waves when they propagate through a strong magnetic or electric field \cite{gertsenshtein1962wave,PhysRevD.16.2915,doi:10.1142/S0217732394003464,PALESSANDRO2023101187}. It was suggested that this effect could be observed near pulsars \cite{PhysRevD.37.1237} and from distinctive anisotropy in cosmic radiation \cite{PhysRevD.49.671}. 

From a general scientific standpoint, there is a compelling interest to test gravitational effects within a controlled laboratory setting, rather than relying solely on the observation of astrophysical phenomena \cite{PhysRevD.15.2047,Bailey2024}. It would enable a more precise comparison between experimental and theoretical results, thus enhancing our confidence in the detection process and in the theories that describe them. However, realizing this feat necessitates the most extreme conditions created in laboratories, with extremely large energy densities involved. Owing to the smallness of Einstein's gravitational constant ($\kappa = 8\pi G/c^{4} \approx 2.1 \times 10^{-43}\; \mbox{m/J}$, where $G$ is the gravitational constant), gravitational effects and more generally, the stretching of space-time, are negligible unless these extreme energy conditions are met. 

With recent advances in laser technologies, which enable unprecedented energy and intensity levels \cite{Bahk:04,Yoon:21,Danson_al_2019}, several theoretical proposals have been put forward to produce LIGE and probe the classical gravity-electromagnetism coupling directly via strong laser fields \cite{PhysRev.37.602,PhysRevD.11.2685, PhysRevD.19.3582,MALLETT2000214,10.1063/1.2169320,ji2006gravitational,Ratzel_2016,Schneiter_2018,PhysRevD.105.104052,morozov2021generation,Spengler_2022,PhysRevD.110.044023} or indirectly through laser-matter interactions \cite{ribeyre2012high,10.1063/1.4962520,kadlecova2017gravitational}. In typical scenarios, the electromagnetic radiation is concentrated at the focal point of a laser beam and can reach relatively high energy densities (typically ranging from $10^{18}$ -- $10^{20}$~J/m$^3$ at the focus for a $10$ -- $100$~J laser at 800 nm focused to the diffraction limit). Owing to the space-time dependence of the resulting electromagnetic radiation, gravitational waves can also be created in this configuration \cite{PhysRevD.110.044023}. 
However, most proposals in the literature predict extremely small metric perturbations for realistic laser parameters, in the range $\| h_{\mu \nu} \| \sim 10^{-40}$--$10^{-35}$ \cite{PhysRevD.105.104052,PhysRevD.110.044023}. Furthermore, it has recently been argued that the existence of a \textit{kugelblitz}---a black hole produced by light---is not plausible due to the Schwinger mechanism at high field strengths, which disperses electromagnetic energy \cite{PhysRevLett.133.041401}. These findings raise the question of whether LIGE generated in laboratories could be experimentally detected. More specifically, are existing or projected experimental methods accurate enough to detect them, and are there other competing quantum electrodynamics (QED) vacuum effects that could hinder their observation? 

\begin{figure}
	\includegraphics[width=0.5\textwidth]{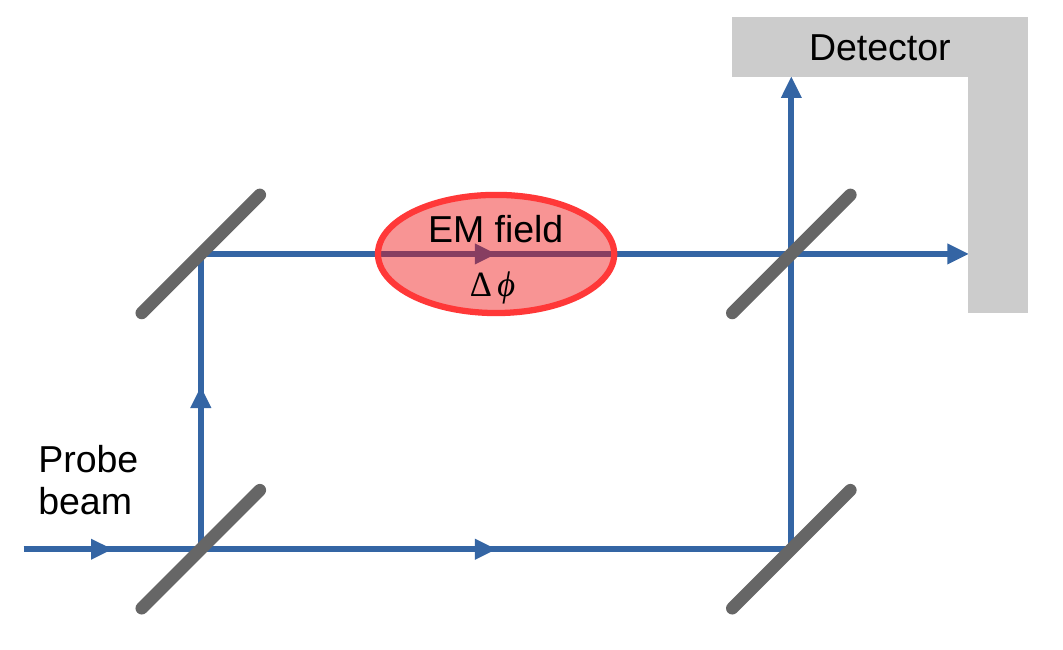}
	\caption{Proposed detection concept of LIGE and QED effects using a MZ interferometer.}
	\label{fig:mz_inter}
\end{figure}

This article is an attempt at answering these questions, in a concrete detection scenario where LIGE are probed via an interferometric technique. The proposed detection scheme, which consists of a Mach-Zehnder (MZ) interferometer, is depicted in Fig. \ref{fig:mz_inter}. It is conceptually the same as the one presented in Ref. \cite{PhysRevA.97.063811} for the detection of vacuum birefringence, but is now applied to LIGE. Other experimental strategies to measure QED vacuum effects are explored in Ref. \cite{Ahmadiniaz_2025}.  Interferometry is one of the most accurate and sensitive technique for the study of gravitational effects and its success is at the basis of gravitational wave detection at LIGO \cite{PhysRevLett.116.061102,PhysRevLett.116.241103}. Furthermore, there exists theoretical bounds on the phase sensitivity of this technique obtained using quantum information theory \cite{DEMKOWICZDOBRZANSKI2015345}, which will allow us to obtain explicit conditions for the observability of LIGE. The MZ interferometer is preferred over other configurations (Michelson, Sagnac and others) because its geometry ensures that photons propagate along distinct, unidirectional paths, thereby simplifying the theoretical analysis as it involves only a single pass of the photons through each arm. In addition, the bounds on phase sensitivity have a simpler form for MZ \cite{DEMKOWICZDOBRZANSKI2015345}. These are the main reasons the MZ interferometer scheme is chosen, although other types of interferometers could also be considered in a similar way. 

This article is separated as follows. In Section \ref{sec:lin_grav}, the formalism of linearized gravity coupled to electromagnetism is presented. Then, in Sections \ref{sec:electric_e} and \ref{sec:electric_e_harm}, we consider the simplest possible scenarios of constant and harmonic homogeneous fields to obtain order-of-magnitude estimates, in the same spirit as Ref. \cite{PhysRevLett.133.041401}. In Section \ref{sec:e_dipole}, a more realistic configuration using an e-dipole electromagnetic field is assessed. We conclude in Section \ref{sec:conclu}.

\section{Linearized gravity coupled to electromagnetism \label{sec:lin_grav}}

The starting point of our analysis is obviously Einstein's equation, which relates the curvature of space-time to the matter-energy content. It is given by \cite{Misner:1973prb,Weinberg:1972kfs,Carroll:2004st}
\begin{align}
	\label{eq:eins}
	R_{\mu \nu} - \frac{1}{2}R g_{\mu \nu} = \kappa T_{\mu \nu},
\end{align}
where $R_{\mu \nu}$ is the Ricci tensor, $R$ is the Ricci scalar, $T_{\mu \nu}$ is the energy-momentum tensor and $\kappa = 8\pi G/c^{4} \approx 2.1 \times 10^{-43}\; \mbox{m/J}$ is the graviton-photon coupling. In the regime where gravitational effects are weak, the metric $g_{\mu \nu}$ is close to the flat Minkowski metric $\eta_{\mu \nu} = \mathrm{diag}[-1,1,1,1]$ while the space-time curvature remains small. In this case, Einstein's equation can be linearized by assuming the following ansatz for the metric:
\begin{align}
	g_{\mu \nu} = \eta_{\mu \nu} + h_{\mu \nu}, \quad \|h_{\mu \nu} \| \ll 1.
\end{align}
Thus, $h_{\mu \nu}$ corresponds physically to a small perturbation over the flat space metric. Using this ansatz, along with the Lorentz gauge condition
\begin{align}
	\partial_{\mu} \bar{h}^{\mu \nu} = 0,
\end{align}
where 
\begin{align}
	\bar{h}_{\mu \nu} = h_{\mu \nu} - \frac{1}{2} \eta_{\mu \nu} h_{\alpha}^{\alpha},
\end{align}
is the trace-reversed metric perturbation, Einstein's equation \eqref{eq:eins} becomes \cite{maggiore_book}
\begin{align}
	\label{eq:wave_eq}
	\left[\frac{1}{c^{2}} \partial_{tt} - \nabla^{2} \right]\bar{h}_{\mu \nu} =  2 \kappa  T_{\mu \nu},
\end{align}
at leading order $O(h)$. This equation describes linearized gravity and is at the heart of our understanding of many weak gravitational phenomena such as gravitational waves \cite{maggiore_book}.

In this regime of small perturbation, Einstein's equation becomes a standard non-homogeneous wave equation sourced by the stress-energy tensor in flat space. An integral representation for the solution of these equations can be obtained once initial conditions are provided. 
We assume that for times $t<0$, all components of the energy-momentum tensor are zero ($T_{\mu \nu} = 0$) and the space is flat. Then, our initial conditions are  $\left. \bar{h}_{\mu \nu}(t)\right|_{t=0} = \left. \partial_{t} \bar{h}_{\mu \nu}(t)\right|_{t=0} = 0$ and Eq. \eqref{eq:wave_eq} has an explicit solution given by \cite{evans2022partial,polyanin2001handbook}
\begin{align}
	\label{eq:h_mu_nu}
	\bar{h}_{\mu \nu}(t,\boldsymbol{x}) &= \frac{\kappa}{2\pi } \int_{\mathcal{B}(ct,\boldsymbol{x})} \frac{T_{\mu \nu} \left(\boldsymbol{y}, t  - \frac{|\boldsymbol{y} - \boldsymbol{x}|}{c} \right)}{|\boldsymbol{y} - \boldsymbol{x}|} d^{3} \boldsymbol{y},
\end{align}
where $\mathcal{B}(ct,\boldsymbol{x})$ is the ball of radius $ct$ centered at $\boldsymbol{x} \in \mathbb{R}^{3}$. This last equation is valid at tree level within a semi-classical approach. Extensions of this formalism that include quantum loop corrections are discussed in Refs. \cite{Fiorenzo_Bastianelli_2005,Bastianelli_2008}.

As we are interested by LIGE, we now describe the electromagnetic energy-momentum tensor. Its contravariant tensor components are given by (for $i,j,k = 1,2,3$)  
\begin{align}
	\label{eq:T_00}
	T^{00} &= \rho ,\\
	\label{eq:T_0i}
	T^{0i} = T^{i0} &=  \frac{1}{c}S^{i} , \\
	\label{eq:T_ij}
	T^{ij} &= - \sigma^{ij},
\end{align} 
where the energy density $\rho$, the Poynting vector $S$ and the Maxwell stress tensor $\sigma$ have been introduced. They are defined as
\begin{align}
	\rho &= \frac{1}{2} \left[ \epsilon_{0} \boldsymbol{E}^{2} +  \frac{1}{\mu_{0}}\boldsymbol{B}^{2}\right], \\
	S^{i} &= \frac{1}{\mu_{0}}\epsilon^{ijk} E^{j} B^{k}, \\
	\sigma^{ij} &=\epsilon_{0} E^{i}E^{j} +  \frac{1}{\mu_{0}}B^{i} B^{j} - \rho \delta^{ij},
\end{align}
where $\boldsymbol{E}$ and $\boldsymbol{B}$ are the electric and magnetic fields, respectively, while $\epsilon^{ijk}$ is the antisymmetric Levi-Civita tensor.

\section{Homogeneous electric field \label{sec:electric_e}}

In this section, the feasibility of observing LIGE via the interferometric technique depicted in Fig. \ref{fig:mz_inter} is assessed for the simplest field configuration: an homogeneous electric field. With this we can obtain an order-of-magnitude estimate of LIGE and understand the physical processes at play. A more physically relevant scenario using a time- and space-dependent laser pulse will be considered in Section \ref{sec:e_dipole}. 

\subsection{Evaluating the metric perturbation}

The electromagnetic field corresponding to an homogeneous electric field is
\begin{align}
	\boldsymbol{E}(t) &= \hat{\boldsymbol{e}}_{x} E_{0}H(t), \\
	\boldsymbol{B}(t) &= 0,
\end{align}
where $E_{0}$ is the field strength, $\hat{\boldsymbol{e}}_{x}$ is the unit vector in direction $x$ and $H(t)$ is the Heaviside step function that turns the field on at $t=0$. 
Then, the energy-momentum tensor becomes
\begin{align}
	T_{\mu \nu} &= \frac{1}{2}\epsilon_{0} E_{0}^{2}\Theta_{\mu \nu} H(t), \\
	\Theta_{\mu \nu} &= 
	\begin{bmatrix}
		1 & 0 & 0 & 0 \\
		0 & -1 & 0 & 0 \\
		0 & 0 & 1 & 0 \\
		0 & 0 & 0& 1
	\end{bmatrix}.
\end{align}
Reporting this into the solution of linearized gravity Eq. \eqref{eq:h_mu_nu} and using translation invariance of the system, we get
\begin{align}
	\bar{h}_{\mu \nu}(t,\boldsymbol{x}) &= \frac{\kappa \epsilon_{0} E_{0}^{2} \Theta_{\mu \nu}}{4\pi } \int_{\mathcal{B}(ct,0)} H \left( t  - \frac{|\boldsymbol{y} |}{c} \right)\frac{d^{3} \boldsymbol{y}}{|\boldsymbol{y}|} , \\
	&= \frac{\kappa \epsilon_{0} E_{0}^{2} \Theta_{\mu \nu}}{4\pi } \int_{0}^{ct} \int_{0}^{2\pi} \int_{0}^{\pi} r \sin(\theta) d\theta d\phi dr , \\
	\label{eq:metric_pertub}
	&= \mathcal{H} \Theta_{\mu \nu} t^{2},
\end{align}
where $ \mathcal{H} = \frac{1}{2} c^{2} \kappa \epsilon_{0} E_{0}^{2} $.
Then, using 
\begin{align}
	h_{\mu \nu} = \bar{h}_{\mu \nu} - \frac{1}{2} \eta_{\mu \nu} \bar{h}_{\alpha}^{\alpha},
\end{align}
and the fact that $\bar{h}_{\alpha}^{\alpha} = 0$, we get $h_{\mu \nu} = \bar{h}_{\mu \nu}$. 

This metric has an intrinsic curvature, as can be demonstrated by evaluating the Ricci scalar:
\begin{align}
	R &= \partial_{\mu} \partial_{\nu} h^{\mu \nu} - \Box h^{\alpha}_{\alpha} ,\\
	\label{eq:ricci_curv}
	  &= \partial_{0}\partial_{0} h^{00} = \kappa \epsilon_{0} E_{0}^{2} . 
\end{align}
This static curvature exists wherever the electric field is non-zero and thus, is expected to have an effect on photon propagation in this region.

\subsection{Induced phase difference in the Mach-Zehnder interferometer}

To estimate the phase difference in the MZ interferometer due to  space-time perturbations induced by the electric field, we use the same method presented in Ref. \cite{maggiore_book}, where an effective time delay is evaluated from the space-time interval. To be more specific, we consider a probe photon propagating through the interferometer (see Fig. \ref{fig:mz_inter}) and reaching the region where the electric field is applied along the second arm of the interferometer. We select a coordinate system where the photon travels along the $z$-coordinate and passes by the point $(t,\boldsymbol{x}) = (0,0)$ when the field is turned-on. Then, it traverses the field region over a distance $L$, the field is turned-off and the photon experienced a delay. 

In this coordinate system, the position of photon emission and measurement (essentially the beginning and end of its propagation in the field) are not modified, even when the field is activated and space-time is curved. This occurs because the coordinate system stretches with the space-time deformation such that the position of the emission and measurement are still $z=0$ and $z=L$, respectively, before and after the field is turned-on at $t=0$. This is very similar to what happens for gravitational waves in transverse-traceless-gauge \cite{maggiore_book}. We demonstrate this property in Appendix \ref{app:test_masses} by showing that test masses at rest remain at rest.

Then, the evaluation of the light-like interval along the photon path is relatively straightforward. It is defined as
\begin{align}
	ds^{2} &= g_{\mu \nu} dx^{\mu} dx^{\nu} = 0, \\
	&=(-1 + \mathcal{H}t^{2}) c^{2} dt^{2} + (1+\mathcal{H} t^{2}) dz^{2},
\end{align}
which yields
\begin{align}
	\label{eq:diff_interval}
	\frac{dz}{dt} &= c  \sqrt{\frac{1 - \mathcal{H} t^{2}}{1 + \mathcal{H}t^{2}}} .
\end{align}
Here, the positive solution was chosen because we are assuming that the probe photons will propagate along the $+z$-axis.
This differential equation can be solved analytically, but it is more simple and convenient to use the fact that the secular term obeys $\mathcal{H}t^{2} \ll 1$ in linearized gravity (for short-enough times as considered in this article). Then, we can expand the right-hand-side of Eq. \eqref{eq:diff_interval} and get
\begin{align}
	\frac{dz}{dt} & = c  (1 - \mathcal{H} t^{2}) + O(h^{2}).
\end{align}
Assuming probe photons propagate between two points separated by a distance $L$ in the coordinate system, we can integrate the previous equation:
\begin{align}
	L &= c \Delta t - c \mathcal{H} \int_{0}^{\Delta t} dt' t'^{2}, \\
	\label{eq:L}
	  &= c \Delta t\left(1 - \frac{\mathcal{H} \Delta t^{2}}{3}\right).
\end{align}
Iterating Eq. \eqref{eq:L} and keeping only terms of $O(\mathcal{H}t^{2})$ yields
\begin{align}
	\label{eq:cdt}
	c \Delta t := L_{\mathrm{eff}} = L + \frac{\mathcal{H}}{3} \frac{L^{3}}{c^{2}} .
\end{align}
The first term in Eq. \eqref{eq:cdt} represents the propagation length in flat space while the second term gives the LIGE correction. This effective length modification could be measured using interferometric methods, which are sensitive to optical path differences. The corresponding phase difference produced by LIGE is thus given by
\begin{align}
	\label{eq:dphi_lige}
	\Delta \varphi_{\mathrm{lige}} &= \frac{2\pi}{\lambda_{\mathrm{probe}}}(L_{\mathrm{eff}} - L) \\
	\label{eq:dphi_lige2}
	&= \frac{\pi \kappa \epsilon_{0} E_{0}^{2} L^{3}}{3 \lambda_{\mathrm{probe}}} = \frac{\kappa U}{2 \lambda_{\mathrm{probe}}},
\end{align}
where $\lambda_{\mathrm{probe}}$ is the wavelength of the probe photon.
This is proportional to the electromagnetic energy $U = \frac{2}{3} \pi \epsilon_{0} E_{0}^{2} L^{3}$ contained in the spherical region of radius $L$ through which the photon is propagating.

\subsection{Conditions for observing LIGE}

The feasibility of detecting LIGE using the MZ interferometer is now discussed. Non-classical light probes can be used to enhance the performance and sensitivity of interferometry. By using coherent and squeezed states of light as inputs to the MZ interferometer, along with a quantum phase estimation protocol, one can reach sensitivities beyond the standard shot-noise limit given by $\Delta \varphi_{\mathrm{shot-noise}} = 1/\sqrt{\bar{N}}$, where $\bar{N}$ is the average number of photons. The most sensitive schemes for the MZ interferometer can theoretically reach the \textit{Heisenberg scaling} for which \cite{PhysRevLett.100.073601,DEMKOWICZDOBRZANSKI2015345}
\begin{align}
	\Delta \varphi_{\mathrm{th}} = \frac{1}{\bar{N}}.
\end{align} 
A concrete implementation of such quantum interferometric techniques have been realized in state-of-the-art gravitational wave detectors like LIGO, for which the experimental phase resolution reaches $\Delta \varphi_{\mathrm{exp}} \approx 4.0 \times 10^{-12}$ radians on the frequency range $[100 \mbox{ Hz},400 \mbox{ Hz}]$, for a laser wavelength of $\lambda = 1064$ nm \cite{PhysRevD.102.062003}. 

To determine the first bound on LIGE detectability, it is required that
\begin{align}
	\label{eq:phi_cond01}
	\Delta \varphi_{\mathrm{lige}} > \Delta \varphi_{\mathrm{th}} .
\end{align}
The average number of photons can be estimated from the probe laser energy as $\bar{N} \approx U_{\mathrm{probe}}/ \hbar \omega_{\mathrm{probe}}$. Therefore, Eq. \eqref{eq:phi_cond01} becomes
\begin{align}
	 U U_{\mathrm{probe}} &> \frac{4\pi \hbar c}{\kappa}, \\
	  &\gtrsim 1.9 \times 10^{18} \mbox{ J}^{2} .
\end{align}
Interestingly, this condition depends only on the probe and static field energies. This is the first requirement to detect LIGE using interferometry.


 
In precise interferometric measurements, numerous sources of noise and competing mechanisms must be carefully controlled to achieve maximal sensitivity beyond the shot noise and close to the Heisenberg limit. A comprehensive discussion of all potential noise sources arising from the experimental apparatus or the environment is outside the scope of this article. Instead, this work focuses on one specific competing mechanism that is expected to arise whenever photons propagate through a static electric field: light-by-light scattering (LLS) \cite{RevModPhys.78.591,FEDOTOV20231,particles3010005,PhysRevD.101.116019}.

When the magnetic field is zero, the polarization induced by LLS and vacuum QED effects at lowest order is given by \cite{Battesti_2013}:
\begin{align}
	\boldsymbol{P} = 4 c_{\mathrm{qed}} \epsilon_{0}^{2} |\boldsymbol{E}|^{2} \boldsymbol{E},
\end{align}
where the constant $c_{\mathrm{qed}} = \frac{2\alpha^{2}\hbar^{3}}{45 m^{4}c^{5}} \approx 1.67 \times 10^{-30}$ m$^{3}$/J is defined in terms of the fine structure constant $\alpha$ and the electron mass $m$. This polarization is obtained from the leading order contribution to the Euler-Heisenberg low energy effective theory, valid when photon energies obey $\hbar \omega \ll mc^{2}$  and the field strength is $|\boldsymbol{E}| \ll E_{S}$, where $E_{S} \approx 1.32 \times 10^{18}$ V/m is the Schwinger critical field strength \cite{doi:10.1142/9789812775344_0014,doi:10.1142/S2010194512007222}. 
In turn, the field induced polarization modifies the vacuum refractive index as \cite{PhysRevA.63.012107}
\begin{align}
\Delta n = 2 c_{\mathrm{qed}} \epsilon_{0}|\boldsymbol{E}|^{2}.
\end{align}
A wave propagating in a static electric field will thus experience a phase difference given by
\begin{align}
	\label{eq:phi_qed}
	\Delta \varphi_{\mathrm{qed}} = \frac{4\pi L}{\lambda_{\mathrm{probe}}} c_{\mathrm{qed}} \epsilon_{0} E_{0}^{2}.
\end{align}
A similar result was obtained in Ref. \cite{PhysRevA.97.063811}. 

To observe LIGE experimentally, it is required that LIGE dominates over QED effects:
\begin{align}
	\Delta \varphi_{\mathrm{lige}} > \Delta \varphi_{\mathrm{qed}} .
\end{align} 
Using the explicit values in Eqs. \eqref{eq:dphi_lige2} and \eqref{eq:phi_qed}, we get the second bound:
\begin{align}
	L &> \sqrt{\frac{12 c_{\mathrm{qed}}}{\kappa}}, \\
	 & \gtrsim L_{\mathrm{min}} \approx 9.8 \times 10^{6} \mbox{ m}.
\end{align}
This condition only depends on the propagation length $L$ and not on the field strength, owing to the fact that both processes exhibit a quadratic dependence on the electric field strength ($\Delta \varphi_{\mathrm{lige}} , \Delta \varphi_{\mathrm{qed}} \propto E_{0}^{2}$). Therefore, this condition only depends on the ratio of their respective coupling strength $\frac{ c_{\mathrm{qed}}}{\kappa}$. In other words, increasing the field strength does not improve the LIGE signal with respect to the QED noise because the latter is also growing at the same rate. 


To summarize, we obtain two required conditions for the observation of LIGE using a MZ interferometer:
\begin{enumerate}
	\item $ U U_{\mathrm{probe}} \gtrsim 1.9 \times 10^{18} \mbox{ J}^{2}$
	\item $L \gtrsim 9.8 \times 10^{6}$ m
\end{enumerate} 
While the former is related to the theoretical phase resolution in the Heisenberg limit, the latter stems from the same power-law dependence on the electric field of $\Delta \varphi_{\mathrm{lige}}$ and $\Delta \varphi_{\mathrm{qed}}$. In both cases, these bounds cannot be easily circumvented. For the second condition, one possibility would consists of having a plane-wave pump beam co-propagating with the probe photons (with the same linear polarization) instead of a homogeneous field. In this case, the induced QED polarization, which is given by \cite{Battesti_2013}
\begin{align}
	\boldsymbol{P}_{\mathrm{QED}} &= 4 c_{\mathrm{qed}} \epsilon_{0}^{2} (\boldsymbol{E}^{2} - c^{2}\boldsymbol{B}^{2}) \boldsymbol{E} + 14 c_{\mathrm{qed}} \frac{\epsilon_{0}}{\mu_{0}} (\boldsymbol{E} \cdot \boldsymbol{B}) \boldsymbol{B} , 
\end{align}
reduces to $\boldsymbol{P}_{\mathrm{QED}} = 0$ (and similarly for the QED-induced magnetization). Therefore, there is no QED-induced phase difference for plane waves and no ``background noise'' to LIGE in this configuration. However, as demonstrated in Appendix \ref{app:plane_wave}, the length difference induced by LIGE is also zero. Therefore, even if the QED signal is vanishing, using plane-waves does not improve the signal-to-noise ratio.

We can now discuss qualitatively if these requirements could be fulfilled in practice. The most energetic laser in the world right now is at the National Ignition Facility (NIF) \cite{Danson_al_2019} and produces laser pulses with an energy of $U_{\mathrm{NIF}} \approx 2.05 \times 10^{6} $ J \cite{PhysRevLett.132.065102}. This is still 6 orders of magnitude lower than what is required, thus making this measurement with lasers very unlikely in the near future. The second requirement is also stringent because it implies the existence of an electric field over a distance scale approaching the Earth diameter ($\approx 12.7 \times 10^{6}$ m), which would be challenging to implement in a laboratory settings. However, it may be performed in a MZ interferometer by using a resonant optical cavity on the interferometer arms. This causes the probe photon to traverse the field regions multiple times, effectively extending the arm lengths. This strategy was employed at LIGO to detect gravitational waves \cite{PhysRevLett.116.061102}.

To conclude this section, some approximations used in obtaining these order of magnitude estimates are critically assessed. 
First, we verify that for all considered conditions, linearized gravity in which $\| h_{\mu \nu} \| \ll 1$, is a good approximation. This is guaranteed provided that $\mathcal{H}t^{2} \ll 1$. The largest electric field considered in the calculation is $E_{0} = E_{S}$ while the largest time scale is $t \sim L_{\mathrm{min}}/c$, leading to $\mathcal{H}t^{2} \approx 1.5 \times 10^{-4} \ll 1$, which is clearly in the linear regime. 

Second, we examine whether restricting the electric field to a finite domain instead of the assumed infinite one, significantly impacts the final result, a finite domain being a more realistic model of a laser field. Assuming the electric field has a support in a subdomain $\Omega$ (i.e. $|\boldsymbol{E}| = E_{0}$ when $\boldsymbol{x} \in \Omega$ and $|\boldsymbol{E}| = 0$ when $\boldsymbol{x} \notin \Omega$) and that we are interested to evaluate the metric perturbation on a set of positions $\boldsymbol{x} \in \Omega_{\boldsymbol{x}} \subset \mathbb{R}^{3}$, the solution in Eq. \eqref{eq:metric_pertub} is valid when every ball defined in Eq. \eqref{eq:h_mu_nu} is fully embedded in the electric field support at final time $t_{f}$. More precisely, we require $\mathcal{B}(c t_{f}, \boldsymbol{x}) \subset \Omega$ for all $\boldsymbol{x} \in \Omega_{\boldsymbol{x}}$. Therefore, as long as the final times are short-enough such that this condition is fulfilled, the obtained results are not impacted by the shape or the extent of $\Omega$.  


Third, the time-dependence of the electric field could potentially have a strong influence on LIGE. While a laser field typically has a harmonic time dependence, our model only considered an electric field turning on at $t=0$. This is investigated further in the next section.

\section{Homogeneous electric field with harmonic time-dependence \label{sec:electric_e_harm}}

The electromagnetic field corresponding to an homogeneous electric field with a harmonic time-dependence is
\begin{align}
	\boldsymbol{E}(t) &= \hat{\boldsymbol{e}}_{x} E_{0}H(t)\sin(\omega t - \phi), \\
	\boldsymbol{B}(t) &= 0,
\end{align}
where $\omega$ is the angular frequency of oscillation and $\phi$ an arbitrary phase. 
Following the same procedure as in Section \ref{sec:electric_e} by evaluating the energy-momentum tensor and reporting its expression in Eq. \eqref{eq:h_mu_nu} while using translation invariance of the system, we get
\begin{align}
	\bar{h}_{\mu \nu}(t,\boldsymbol{x}) 
	&= \kappa \epsilon_{0} E_{0}^{2} \Theta_{\mu \nu} \int_{0}^{ct}  r \sin^{2} \left[\omega \left(t - \frac{r}{c}\right) - \phi \right]  dr , \\
	\label{eq:metric_pertub_harm}
	&= \mathcal{H} \Theta_{\mu \nu}  \biggl[ \frac{t^{2}}{2} - \frac{t \sin(2\phi)}{2\omega} \nonumber \\
	& \quad \quad \quad  + \frac{\cos(2\omega t - 2\phi) - \cos(2\phi)}{4 \omega^{2}}\biggr]. 
\end{align}
The Ricci scalar now becomes
\begin{align}
	R &= \partial_{0}\partial_{0} h^{00} = \frac{\kappa \epsilon_{0} E_{0}^{2}}{2} \bigl[ 1 -\cos(2\omega t - 2\phi) \bigr] .
\end{align}
The curvature is now oscillating in time, at twice the frequency of the electric field. Again, this curvature should have an effect on travelling probe photons. As the phase does not change the magnitude of the curvature, it is now set to $\phi = 0$.

Then, the evaluation of the light-like interval along the photon path proceeds as for the homogeneous field case. This yields 
\begin{align}
	\frac{dz}{dt} & = c  \left\{ 1 - \mathcal{H} \left[ \frac{t^{2}}{2} + \frac{\cos(2\omega t) - 1}{4 \omega^{2}}\right] \right\} + O(h^{2}).
\end{align}
Assuming probe photons propagate between two points separated by a distance $L$ in the coordinate system, we can integrate the previous equation:
\begin{align}
	L &= c \Delta t - c \mathcal{H} \int_{0}^{\Delta t} dt' \left[ \frac{t^{2}}{2} + \frac{\cos(2\omega t) - 1}{4 \omega^{2}}\right], \\
	\label{eq:L_harm}
	&= c\Delta t \biggl\{ 1 -  \mathcal{H} \left[\frac{\Delta t^{2}}{6}  + \frac{1}{4 \omega^{2}} \biggl(\mathrm{sinc}(2\omega \Delta t) - 1 \biggr) \right] \biggr\} .
\end{align}
Iterating Eq. \eqref{eq:L_harm} and keeping only terms of $O(\mathcal{H})$ yields
\begin{align}
	\label{eq:cdt_harm}
	c \Delta t &:= L_{\mathrm{eff}} =   L +    \mathcal{H} \left[\frac{L^{3}}{6c^{2}}  + \frac{L}{4 \omega^{2}} \biggl[\mathrm{sinc}\left(2\omega \frac{L}{c}\right) - 1 \biggr] \right]   .
\end{align}
Again, the first term on the right-hand-side of Eq. \eqref{eq:cdt_harm} represents the propagation length in flat space while the next terms give the LIGE correction. Furthermore, because $\mathrm{sinc}\left(2\omega \frac{L}{c}\right) - 1  \leq 0$, the induced phase difference for the time-harmonic field is always smaller than for the constant field: $\Delta \varphi_{\mathrm{lige}}^{\mathrm{(harmonic)}} < \Delta \varphi_{\mathrm{lige}}$. Therefore, the electric field time-dependence effectively reduces LIGE and thus, the bounds obtained in the previous section should be interpreted as minimal requirements --- for a given electric field strength, the time-dependence does not improve the detection feasibility, but rather makes it more challenging.

%

\section{Quasi-Gaussian e-dipole model \label{sec:e_dipole}}

In previous sections, order-of-magnitude estimates have been obtained in two simple field configurations. While these cases are useful to obtain scaling laws and to understand the physics involved, they do not accurately represent a real laser field. In this section, we consider a more realistic scenario where LIGE is induced by a tightly focused field described by the e-dipole model. We resort to this field configuration for two main reasons: 1) these exact solutions of Maxwell's equation in vacuum are relatively close to realistic tightly-focused configurations and could be generated experimentally by using several counterpropagating focused pulses \cite{Marklund_2023} or by using a large high numerical aperture parabolic mirror with an incident radially polarized beam \cite{PhysRevA.86.053836} and 2) at the focus, they yield a time-dependent pure electric field, reminiscent of the cases considered in previous sections. In this sense, e-dipole pulses represent single-beam solutions that most closely resemble the static and oscillating field configurations considered in Sections \ref{sec:electric_e} and \ref{sec:electric_e_harm}.   
They are often deemed as ``optimal pulses'' because they yield the maximum possible field amplitude for a given incoming power and frequency \cite{PhysRevA.86.053836}. For this reason, they have been used for many applications where high field strengths are required, such as the Schwinger pair production \cite{PhysRevLett.111.060404} and other quantum electrodynamics observables \cite{PhysRevE.99.031201,PhysRevE.105.065202}. However, they are chosen here for their simple form at the focus, rather than for their potential to produce high intensities. As demonstrated in Sec. \ref{sec:electric_e}, higher intensities do not improve the signal-to-noise ratio.
The main goal of this section is to assess whether LIGE from e-dipoles could be detected in the near future and whether they follow similar scaling laws as homogeneous fields. 


\subsection{Field model and effective length difference}

The Gaussian e-dipole model represents a tightly focused laser beam with an optimal field amplitude. It is a smooth exact solution to Maxwell's equation obtained from Hertz vectors. It is given explicitly by \cite{PhysRevA.86.053836}
\begin{align}
	\boldsymbol{E}(t,\boldsymbol{x}) &= \frac{1}{4\pi \epsilon_{0}} \biggl\{
	\frac{\hat{\boldsymbol{x}} \times [\hat{\boldsymbol{x}} \times \boldsymbol{d}_{0}]}{c^{2}r} \ddot{g}_{-}(t,r) \nonumber \\
	&+ \frac{3\hat{\boldsymbol{x}}(\hat{\boldsymbol{x}} \cdot \boldsymbol{d}_{0}) - \boldsymbol{d}_{0}}{r^{3}} \left[ \frac{r}{c} \dot{g}_{+}(t,r) + g_{-}(t,r) \right]
	\biggr\} ,\\
	\boldsymbol{B}(t,\boldsymbol{x}) &= \frac{\mu_{0} c}{4\pi} (\boldsymbol{d}_{0} \times \hat{\boldsymbol{x}}) \left[ \frac{1}{c^{2}r} \ddot{g}_{+}(t,r) + \frac{1}{cr^{2}} \dot{g}_{-}(t,r) \right],
\end{align}
where 
\begin{align}
	g_{\pm}(t,r) = g\left(t - \frac{r}{c}\right) \pm g\left(t + \frac{r}{c}\right),
\end{align}
and where $\boldsymbol{d}_{0}$ is a constant virtual dipole while $g(t)$ is an arbitrary driving function. The dipole essentially sets the magnitude and direction (polarization) of the electric field at the focus. Its normalization and relation to the maximum electric field strength are given in Appendix \ref{app:time_derivatives}. For the quasi-Gaussian e-dipole model, the driving function is chosen as
\begin{align}
	g(t) = e^{-(a t)^{2}} \sin (\omega t),
\end{align}
where $\omega = 2\pi c/\lambda$ is the angular frequency of the laser beam ($\lambda$ is the wavelength) and $a$ gives the pulse duration. For completeness, the time derivatives of the driving function are also given in Appendix \ref{app:time_derivatives}.

The metric perturbation is then evaluated from the expression of the energy-momentum tensor Eqs. \eqref{eq:T_00}--\eqref{eq:T_ij} and the solution to the linearized gravity Eq. \eqref{eq:h_mu_nu}. The integral in Eq. \eqref{eq:h_mu_nu} is evaluated numerically with a high-performance code using a domain decomposition parallelization and an adaptative cubature numerical method based on the Cubature library \cite{cubature}. 

The effective length difference induced by LIGE is evaluated from the photon time delay in the semi-classical approximation, in a slightly more general way than for homogeneous fields. The time delay is calculated by starting from the conditions for null geodesics, which are followed by photons:
\begin{align}
	g_{\mu \nu} \dot{x}^{\mu} \dot{x}^{\nu} = 0.
\end{align}
As in Sec. \ref{sec:electric_e}, we consider a propagation along the $z$-axis only (thus $\dot{x}^{1} = \dot{x}^{2} = 0$) and linearized gravity, so we explicitly get
\begin{align}
	(-1 + h_{00}) (\dot{x}^{0} )^{2} + (1 + h_{33}) (\dot{x}^{3})^{2} + 2 h_{03} \dot{x}^{0} \dot{x}^{3}  = 0.
\end{align}
Dividing by $(\dot{x}^{0} )^{2}$ yields:
\begin{align}
	 (1 + h_{33}) \left[\frac{dz}{c dt} \right]^{2} + 2 h_{03}  \frac{dz}{c dt} + (-1 + h_{00})  = 0,
\end{align}
whose solution is
\begin{align}
	\label{eq:cdt_dz}
	c dt &= \frac{(1+ h_{33})}{-h_{03} \pm \sqrt{ h_{03}^{2} +  (1+h_{33}) (1 - h_{00})}} dz , \\
	&= \left[ 1 + \frac{h_{00} + 2h_{03} + h_{33}}{2} \right] dz + O(h^{2}), \\
	&= \left[1 + \frac{\varkappa }{2} \right] dz, \;\; \mbox{with} \;\; \varkappa = h_{00} + 2h_{03} + h_{33} .
\end{align}
In Eq. \eqref{eq:cdt_dz}, the ``+'' solution is picked because we assume the probe photons propagates in the $+z$-direction with $dz/dt > 0$.
We assume here an initial condition given by $t(z_{0}) = t_{0}$, or in other words, that the photon is emitted at the spacetime coordinate $ (t_{0},0,0,z_{0})$.
The formal solution to this non-autonomous ordinary differential equation gives the time delay of the photon. It is given formally by
\begin{align}
	t(z) = t(z_{0}) + \frac{1}{c} \int_{z_{0}}^{z} \left[1 + \frac{\varkappa(t(z'),z')}{2} \right] dz'.
\end{align}
Then, assuming the photon propagates over a distance $L = z - z_{0}$ and linearizing the solution by iterating the time argument, we obtain
\begin{align}
	\label{eq:dt_dipole}
	c\Delta t 
	&= L   + \Delta L  + O(h^{2}), 
\end{align}
where $\Delta t = t - t_{0}$ and with the effective length difference: 
\begin{align}
	\label{eq:dl}
	\Delta L = \frac{1}{2} \int_{z_{0}}^{z_{0} + L}  \varkappa\left(t_{0} + \frac{z' - z_{0}}{c},z'\right)  d z'.
\end{align}
This expression is equivalent to the one found in Refs. \cite{CIUFOLINI2003101,Ciufolini_2002}: while the first term on the right-hand-side of Eq. \eqref{eq:dt_dipole} represents flat-space propagation, the second term $\Delta L$ can be related to a time delay. The latter is now evaluated numerically.   

\subsection{Numerical results: scaling laws}

To obtain scaling laws and understand how the effective length difference $\Delta L$ varies as a function of the e-dipole model parameters, a systematic numerical study is now performed. This investigation focuses on five important parameters: the maximum field strength, the propagation distance of the probe photons, the pulse duration, the laser wavelength and the electric field polarization at the focus. The numerical values of these parameters are provided in Table \ref{tab:e_dip_params}, which is our ``default setup''. Except for the maximum electric field strength, which is close to the Schwinger critical field $E_{S}$, all parameter values are close to existing high intensity lasers. Then, we subsequently select each of these parameters and perform a parameter sweep while keeping the others fixed. 

In all simulations, the probe photon's trajectory is selected to ensure it arrives at the laser beam's focal point precisely when the electric field reaches it maximum value. Realizing this experimentally is highly challenging, as it demands extremely precise spatio-temporal control of both beams -- with timing accuracy better than the pulse duration and spatial alignment within a fraction of the wavelength. However, there exists techniques that enable coherent superposition of short pulses, which could  be adapted to our setup \cite{Mu:16}.

\begin{figure*}
	\centering
	\begin{subfigure}{0.4\textwidth}
		\includegraphics[width=\textwidth]{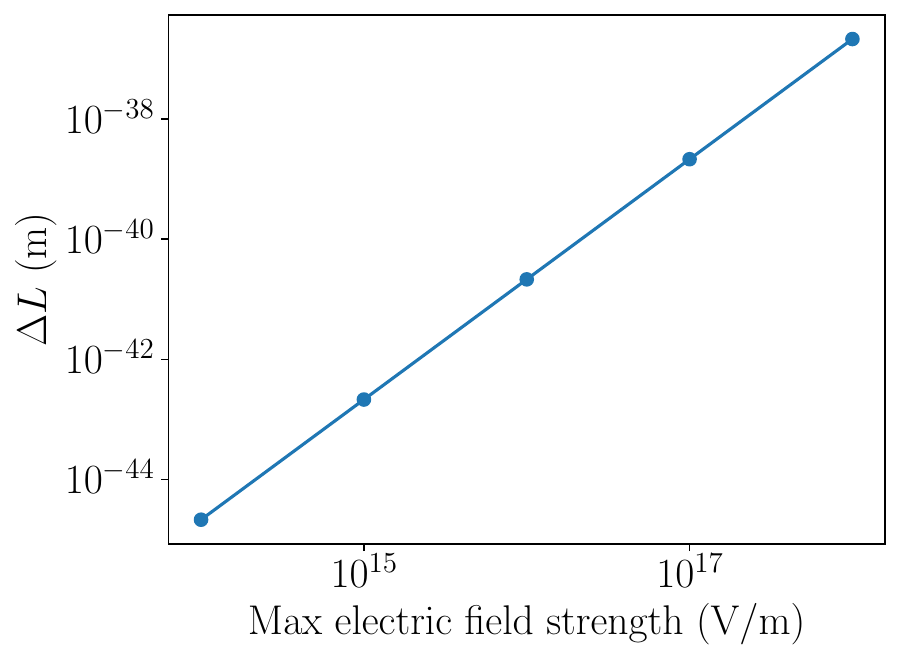}  
		\label{fig:field_strength}
	\end{subfigure}
	%
	\begin{subfigure}{0.4\textwidth}
		\includegraphics[width=\textwidth]{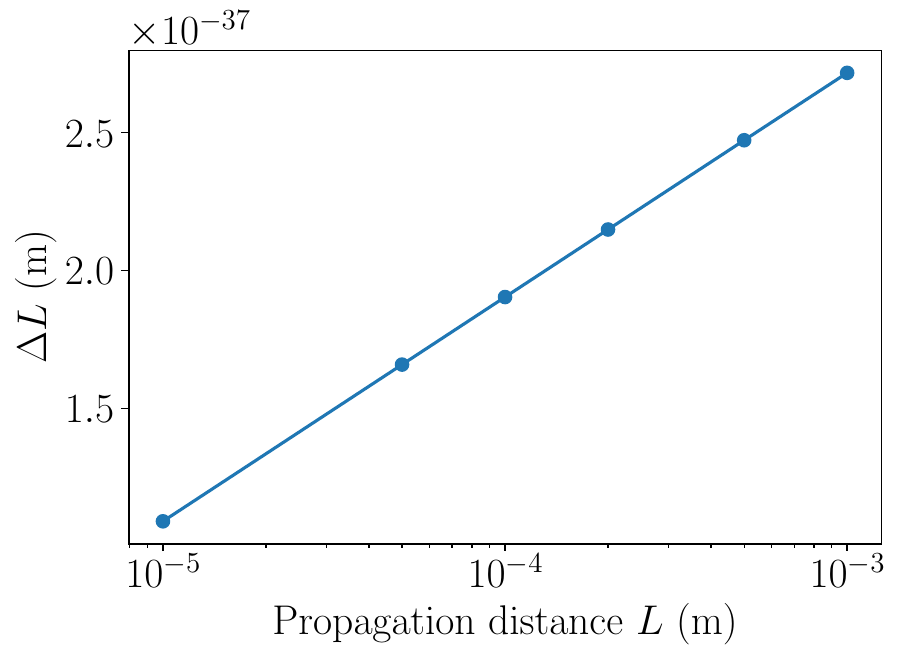}
		\label{fig:prop_dist}
	\end{subfigure}
	%
	\begin{subfigure}{0.4\textwidth}
		\includegraphics[width=\textwidth]{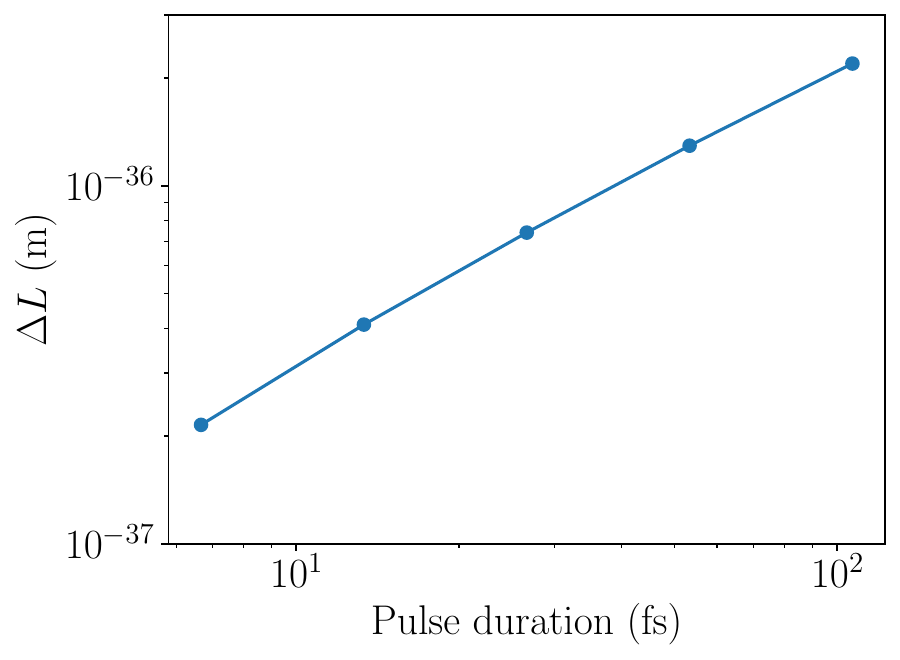}
		\label{fig:pulse_duration}
	\end{subfigure}
	%
	\begin{subfigure}{0.4\textwidth}
		\includegraphics[width=\textwidth]{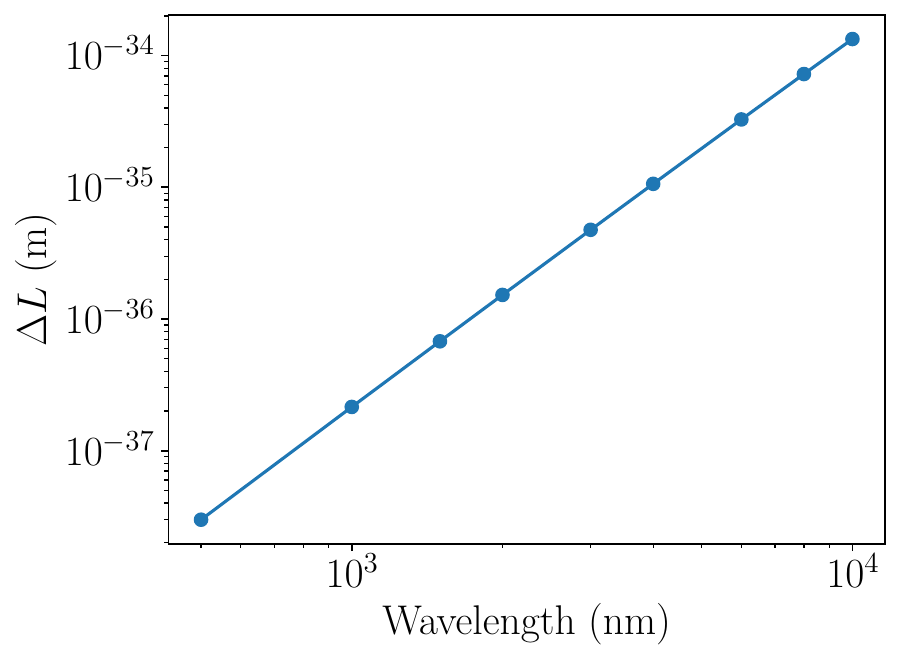}
		\label{fig:prop_wavelength}
	\end{subfigure}
	%
	\caption{Numerical results for the effective length $\Delta L$ dependence on maximum field strength, propagation distance, pulse duration and wavelength. }
	\label{fig:num_results}
\end{figure*}

The length difference is then evaluated from Eq. \eqref{eq:dl} using a simple numerical integration routine based on the midpoint rule. The number of quadrature points was chosen to have convergence of the numerical results. In Eq. \eqref{eq:dl}, the metric perturbation is also obtained numerically, using the technique described in the last section. The numerical results are displayed in Fig. \ref{fig:num_results}.

\begin{table}
	\caption{Parameters of the e-dipole beam and the probe photon for the default setup.}
	\label{tab:e_dip_params}
	\begin{tabular}{l c}
		\hline \hline
		Parameters & Value \\
		\hline 
		Wavelength & 1000 nm \\
		Max electric field strength & $1.0 \times 10^{18}$ V/m \\
		Pulse duration  & 2 cycles \\
		Polarization at the focus & $x$-axis \\
		Propagation length ($L$) & 200 $\mu$m \\
		\hline \hline
	\end{tabular}
\end{table}

The numerical results allow us to extract the following scaling law as a function of the maximum electric field $E_{\mathrm{max}}$, the propagation distance of the probe $L$, the wavelength $\lambda$ and the pulse duration $T$:
\begin{align}
	\label{eq:power_law}
	\frac{\Delta L}{L_{0}} \propto E_{\mathrm{max}}^{2} \left(\frac{L}{L_{0}}\right)^{b_{1}} \left(\frac{\lambda}{L_{0}}\right)^{b_{2}} \left(\frac{T}{T_{0}}\right)^{b_{3}},
\end{align}
where $L_{0} = 1000$ nm is a characteristic length and $T_{0} = 10$ fs is some arbitrary characteristic time. Dividing by the characteristic scales and fitting the data, we find that $b_{1} \approx 0.20$, $b_{2} \approx 2.80$ and $b_{3} \approx 0.84$. While the quadratic dependence on the electric field strength is straightforward to understand --- it comes from the quadratic power dependence of the energy momentum tensor in Eq. \eqref{eq:T_00}-\eqref{eq:T_ij} --- the other power laws need more explanation. 

The exponent $b_{1}$ of $L$ is significantly weaker than that of a homogeneous field, for which $\Delta L \propto L^{3}$. This discrepancy occurs because, in the e-dipole model, the electromagnetic field energy is concentrated in the vicinity of the focal point, within a region of size $\ell_{f} \sim \lambda$. For longer distances, the probe photon also interacts with this small high-field region of fixed size $\ell_{f}$, while the remainder of its propagation occurs through weaker fields. Since weaker fields have a lower effect on $\Delta L$ due to the relationship $\Delta L \propto E_{\mathrm{max}}^{2}$, the dependence of $\Delta L$ on the propagation distance is reduced. This also explains the wavelength dependence ($b_{2}$): the wavelength $\lambda$ sets the size of the high-field region $\ell_{f}$, effectively making for a propagation distance of $L \sim \lambda$ in the region with a dominant contribution to $\Delta L$. Thus, the exponent $b_{2}$ approaches that of the homogeneous case for which $b_{1} = 3$, consistent with our observations. Finally, the exponent $b_{3}$ for the pulse duration is close to one, possibly because the pulse total energy is proportional to $T$. 

We also looked at the effect of polarization at the focus, by choosing $\boldsymbol{d}_{0}$ to be pointing in the $z$-direction, along the propagation of the probe photon. The effective length difference in this case is $\Delta L \approx 1.99 \times 10^{-37}$ m, approximately the same ($\Delta L \approx 2.15 \times 10^{-37}$) as when the dipole is in the $x$-direction. We conclude that LIGE is weakly dependent on the direction of the electric field at the focus.

Therefore, the most interesting configurations are the ones with large electric field strength, large propagation distance, large wavelength and longer pulse duration. This is not surprising as these configurations maximize the amount of energy in the electromagnetic field and the time spent by the probe photon in the high intensity region. Even with the high field strength considered in our investigation, which is several orders of magnitude above realistic laser fields ($\sim 10^{15}$ V/m \cite{Yoon:21}), LIGE remains largely under the detection threshold of the most sensitive interferometric techniques (which can probe length difference of $\Delta L_{\mathrm{exp}} \sim  10^{-19}$ m \cite{PhysRevD.102.062003}). The maximum $\Delta L_{\mathrm{max}} \sim 10^{-34}$ m obtained in our configurations at large wavelength is on the order of the Planck length ($1.6 \times 10^{-35}$ m) and is below this accuracy. 

In principle, pushing the limits of laser technologies to reach higher field strengths would increase $\Delta L$ quadratically and improve our chance of observing LIGE. However, the maximum field strength attainable is constrained by QED effects. In a perfect vacuum, the Schwinger effect is triggered when $E_{\mathrm{max}} \lesssim E_{S}$ and electron-positron pairs are spontaneously generated from the field. When this happens, the electromagnetic field is depleted by QED cascades which involve a sequence of high-energy photons and electron-positron pairs emission via the Breit-Wheeler and Compton processes. The end result is a limitation on the intensity that can be reached by a laser \cite{PhysRevLett.105.080402,PhysRevLett.118.154803}. For this reason, we have not considered field strengths above $E_{S}$ in our numerical calculations. 

To increase the LIGE signal using e-dipole fields, another possibility is to take advantage of longer wavelengths. Assuming that the scaling law in Eq. \eqref{eq:power_law} can be extrapolated outside of its tested range and keeping the ratio $L/\lambda = 200$, the field strength and the pulse duration as in the default configuration, we get $\Delta L = C \left(\lambda/L_{0}\right)^{b_{1} + b_{2}}$, where $C \approx 2.15 \times 10^{-37}$ m. Then, to reach $\Delta L \gtrsim \Delta L_{\mathrm{exp}} = 10^{-19}$ m, wavelengths in the radio frequency range of $\lambda \approx 0.78$ m and higher are required. A similar and more realistic configuration will be evaluated numerically in Section \ref{sec:radar} for a radar source.


\subsection{Electromagnetic radiation sources}


In this section, $\Delta L$ will be evaluated for existing electromagnetic radiation sources to test whether they could be used to study LIGE. Three types of sources will be investigated: pulsed high-power lasers, high-energy lasers and long wavelength radars. While the first two can reach high peak intensities and energies, the latter has much longer wavelengths which could compensate for lower electric field strengths. In all cases, we are assuming a tightly focused beam described by the e-dipole field model.  

\subsubsection{Short-pulse high-power lasers}

We consider a high-power, short-pulse laser. One of the state-of-the-art facilities for such lasers is the Extreme Light Infrastructure - Nuclear Physics (ELI-NP), where a 10 PW laser has been successfully demonstrated \cite{eli_np__2022}. The laser parameters are given in Table \ref{tab:elinp_params}. The maximum electric field strength given in the table is for an e-dipole model and is obtained from Eqs. \eqref{eq:d0_edip} and \eqref{eq:e_field_max} by using the pulse energy and duration. 

\begin{table}
	\caption{Parameters of the e-dipole beam and the probe photon for the 10 PW laser at ELI-NP \cite{eli_np__2022}.}
	\label{tab:elinp_params}
	\begin{tabular}{l c}
		\hline \hline
		Parameters & Value \\
		\hline 
		Wavelength & 800 nm \\
		Pulse duration  & 22 fs \\
		Pulse energy & 300 J \\
		Max electric field strength & $1.6 \times 10^{16}$ V/m \\
		Polarization at the focus & $x$-axis \\
		Propagation length ($L$) & 200 $\mu$m \\
		\hline \hline
	\end{tabular}
\end{table}

The effective length difference $\Delta L$ is evaluated numerically using the parameters given in Table \ref{tab:elinp_params}. We obtain
\begin{align}
	\Delta L_{\mathrm{eli-np}} \approx 1.04 \times 10^{-40} \;\; \mbox{m}.
\end{align}
This value is many orders of magnitude below the sensitivity of any existing interferometric techniques. 

\subsubsection{Pulsed high-energy lasers}

We now consider a pulsed high-energy laser. As mentioned earlier, the laser with the most energy has been developed at NIF for triggering nuclear fusion reactions \cite{PhysRevLett.132.065102}. The laser parameters are given in Table \ref{tab:nif_params}. Again, the maximum electric field strength given in the table is for an e-dipole model and is obtained from Eqs. \eqref{eq:d0_edip} and \eqref{eq:e_field_max} by using the pulse energy and duration.

\begin{table}
	\caption{Parameters of the e-dipole beam and the probe photon for the megajoule laser at NIF \cite{PhysRevLett.132.065102}.}
	\label{tab:nif_params}
	\begin{tabular}{l c}
		\hline \hline
		Parameters & Value \\
		\hline 
		Wavelength & 351 nm \\
		Pulse duration  & 5 ns \\
		Pulse energy & $2.05 \times 10^{6}$ J \\
		Max electric field strength & $6.3 \times 10^{15}$ V/m \\
		Polarization at the focus & $x$-axis \\
		Propagation length ($L$) & 200 $\mu$m \\
		\hline \hline
	\end{tabular}
\end{table}

The effective length difference $\Delta L$ is evaluated numerically using the parameters given in Table \ref{tab:nif_params}. We obtain
\begin{align}
	\Delta L_{\mathrm{nif}} \approx 4.69 \times 10^{-41} \;\; \mbox{m}.
\end{align}
Just like for ELI-NP, this value is many orders of magnitude below the sensitivity of any existing interferometric techniques. Therefore, it is unlikely that LIGE could be detected using tightly focused beams generated by present-day laser technologies.

\subsubsection{High-power pulsed radar \label{sec:radar}}

Finally, we consider a high-power radar capable of generating light pulses at much longer wavelengths than those produced by lasers, specifically in the radio range ( $\lambda \sim [0.01, 10]$ m). According to Eq. \eqref{eq:power_law}, these longer wavelengths increase the LIGE signal significantly. Typical high-power radars can reach close to 50 MW for pulses ranging from 10 ns -- 1 $\mu$s. Choosing optimistic but realistic power (50 MW) and pulse duration (50 ns) yields e-dipole model parameters given in Table \ref{tab:radar_params}. Again, the maximum electric field strength given in the table is obtained from Eqs. \eqref{eq:d0_edip} and \eqref{eq:e_field_max} by using the pulse energy and duration.

\begin{table}
	\caption{Parameters of the e-dipole beam and the probe photon for a high-power pulsed radar}
	\label{tab:radar_params}
	\begin{tabular}{l c}
		\hline \hline
		Parameters & Value \\
		\hline 
		Wavelength & 1 m \\
		Pulse duration  & 50 ns \\
		Pulse energy & 2.5 J \\
		Max electric field strength & $7.7 \times 10^{5}$ V/m \\
		Polarization at the focus & $x$-axis \\
		Propagation length ($L$) & 100 m \\
		\hline \hline
	\end{tabular}
\end{table}

The effective length difference $\Delta L$ is evaluated numerically using the parameters given in Table \ref{tab:radar_params}. We obtain
\begin{align}
	\Delta L_{\mathrm{radar}} \approx 6.15 \times 10^{-43} \;\; \mbox{m}.
\end{align}
Just like for lasers considered previously, this value is many orders of magnitude below the sensitivity of any existing interferometric techniques. Using the scaling law in Eq. \eqref{eq:power_law}, we can estimate the radar power required to reach $\Delta L \gtrsim \Delta L_{\mathrm{exp}} = 10^{-19}$ m, i.e. LIGO's resolution. The power required is on the order of $P \approx 8.1 \times 10^{30}$~W. Again, it is unlikely that LIGE could be detected using tightly focused beams generated by present-day radar technologies.

\section{Conclusion \label{sec:conclu}}

In this work, a scheme to detect LIGE was introduced, based on a MZ interferometer. Using simple homogeneous field models, a set of stringent bounds was given for the detectability of LIGE using this experimental technique. Then, more realistic field configurations using the e-dipole model were considered to obtain scaling laws and to investigate the possibility of observing LIGE at existing laser infrastructures and using high-power radars. In all cases, it was demonstrated that detecting LIGE would be a very challenging task, which necessitates major improvements in interferometric techniques and laser field energies and intensities. Also, it was shown that QED effects would always be a competing mechanism to LIGE, which could be surpassed by having long probe propagation lengths. 

Obviously, this work did not consider all possible field configurations. Nevertheless, the constant homogeneous field discussed in Section \ref{sec:electric_e} is possibly close to the optimal configuration. Indeed, we have noticed that adding time- and space-dependence made the effective length $\Delta L$ smaller, for a given field strength. 

Other possible avenues of study include these other field configurations, but also other detection schemes. In this work, we have focused on interferometry because it was successful in the detection of gravitational waves, a weak gravitation effect. However, there exists other techniques and some of them have been considered in Ref. \cite{Spengler_2022}. Determining whether these alternative techniques provide more promising prospects for detecting LIGE remains an open problem.

\begin{acknowledgments}
	The authors would like to thank J.-S. Gagnon for discussions on this subject. This research was enabled in part by support provided by Calcul Qu\'{e}bec (\url{www.calculquebec.ca}) and the Digital Research Alliance of Canada (\url{alliancecan.ca}).
\end{acknowledgments}

\appendix

\section{Test masses initially at rest remain at rest \label{app:test_masses}}

Following Ref. \cite{maggiore_book}, it is now demonstrated that test masses initially at rest in our coordinate system remain at rest at all subsequent times. The starting point is the geodesic equation evaluated at an initial proper time $\tau = 0$. This is given by
\begin{align}
	\left. \frac{d^{2}x^{i}}{d \tau^{2}} \right|_{\tau = 0} &= - \left[\Gamma_{\nu \rho}^{i} \frac{dx^{\nu}}{d \tau} \frac{dx^{\rho}}{d \tau}  \right]_{\tau = 0},
\end{align}
where $\Gamma_{\nu \rho}^{\mu}$ is the Christoffel symbol.
Assuming the mass particle is initially at rest yields $dx^{i}/d\tau |_{\tau=0} = 0$, yielding
\begin{align}
	\left. \frac{d^{2}x^{i}}{d \tau^{2}} \right|_{\tau = 0} &= - \left[\Gamma_{00}^{i} \left(\frac{dx^{0}}{d \tau} \right)^{2}   \right]_{\tau = 0}.
\end{align}
However, we have
\begin{align}
	\Gamma^{i}_{00} = \frac{1}{2} (2 \partial_{0} h_{0i} - \partial_{i} h_{00}) = 0,
\end{align}
because according to Eq. \eqref{eq:metric_pertub}, we have $h_{0i} = \partial_{i} h_{00} = 0$ and thus
\begin{align}
	\left. \frac{d^{2}x^{i}}{d \tau^{2}} \right|_{\tau = 0} &= 0,
\end{align}
implying that $dx^{i}/d \tau$ is zero at all times.
Therefore, similar to the transverse-traceless (TT) frame, our frame stretches itself when the electric field is turned on, such that the position of free test masses initially at rest does not change. In other words, freely falling objects mark the positions in the coordinate system. 



\section{Plane-wave electromagnetic field \label{app:plane_wave}}

The plane-wave electromagnetic field propagating in the $+z$-direction is given by
\begin{align}
	\boldsymbol{E}(t,z) &= \hat{\boldsymbol{e}}_{x} E_{0}H(t)\cos(\eta), \\
	\boldsymbol{B}(t,z) &= \hat{\boldsymbol{e}}_{y} \frac{E_{0}}{c}H(t)\cos(\eta),
\end{align}
where $\eta = \omega t - kz$ and where $k=\omega/c$ is the wavenumber. Then, the energy-momentum tensor becomes
\begin{align}
	T_{\mu \nu} &= \frac{1}{2}\epsilon_{0} E_{0}^{2}\Theta'_{\mu \nu} H(t)\left[1 + \cos(2\eta)\right], \\
	\Theta'_{\mu \nu} &= 
	\begin{bmatrix}
		1 & 0 & 0 & -1 \\
		0 & 0 & 0 & 0 \\
		0 & 0 & 0 & 0 \\
		-1 & 0 & 0& 1
	\end{bmatrix}.
\end{align}
Reporting this expression into Eq. \eqref{eq:h_mu_nu} and performing the variable change $\boldsymbol{z} = \boldsymbol{y} - \boldsymbol{x}$, we get
\begin{align}
	\bar{h}_{\mu \nu}(t,\boldsymbol{x}) &= \frac{\kappa \epsilon_{0}E_{0}^{2}}{4\pi} \Theta'_{\mu \nu}  \int_{\mathcal{B}(ct,0)} \frac{1}{|\boldsymbol{z}|}\biggl[1 +  \nonumber \\
	&+ \frac{e^{2i \left[\omega \left(t - \frac{|\boldsymbol{z}|}{c} \right) - k(z_{3} + x_{3})\right]} }{2}
	\nonumber \\
	&+ \frac{ e^{-2i \left[\omega \left(t - \frac{|\boldsymbol{z}|}{c} \right) - k(z_{3} + x_{3})\right]}}{2}\biggr] d^{3} \boldsymbol{z}.
\end{align}
Switching to spherical coordinates, we get
\begin{align}
	\bar{h}_{\mu \nu}(t,\boldsymbol{x}) &= \frac{\kappa \epsilon_{0}E_{0}^{2}}{2} \Theta'_{\mu \nu} \nonumber \\
	&\times  \biggl\{ \int_{0}^{ct}dr \int_{0}^{\pi} d\theta r \sin(\theta)  \nonumber \\
	&+e^{2i\eta}\int_{0}^{ct}dr \int_{0}^{\pi} d\theta r \sin(\theta) \frac{e^{-2ikr \left[1 + \cos \theta\right]} }{2} \nonumber \\
	&+e^{-2i\eta}\int_{0}^{ct}dr \int_{0}^{\pi} d\theta r \sin(\theta) \frac{ e^{2ikr \left[1 + \cos \theta \right]}}{2} \biggr\}.
\end{align}
Then, we use the following known integral:
\begin{align}
	I &= \int_{0}^{\pi} \sin \theta e^{ \pm i b \cos \theta} d\theta, \\
	  &= 2\frac{\sin b}{b},
\end{align}
to get
\begin{align}
	\bar{h}_{\mu \nu}(t,\boldsymbol{x}) &= \frac{\kappa \epsilon_{0}E_{0}^{2}}{2} \Theta'_{\mu \nu}   \biggl\{ c^{2} t^{2}  \nonumber \\
	&+e^{2i\eta}\int_{0}^{ct}dr   \sin(2kr) \frac{e^{-2ikr } }{2k} \nonumber \\
	&+e^{-2i\eta}\int_{0}^{ct}dr  \sin(2kr) \frac{ e^{2ikr}}{2k} \biggr\}.
\end{align}
The last integral can also be performed analytically, and we get
\begin{align}
	\label{eq:h_munu_pw}
	\bar{h}_{\mu \nu}(t,\boldsymbol{x}) &= \frac{\kappa \epsilon_{0}E_{0}^{2}}{2} \Theta'_{\mu \nu}   \biggl\{ c^{2} t^{2}  \nonumber \\
	&+\frac{e^{2i\eta}}{2k}\left[\frac{1}{8k} - \frac{ic t}{2} - \frac{e^{-4i \omega t}}{8k}\right] \nonumber \\
	&+\frac{e^{-2i\eta}}{2k}\left[\frac{1}{8k} + \frac{ic t}{2} - \frac{e^{4i \omega t}}{8k}\right] \biggr\}.
\end{align}
Finally, combining everything together yields
\begin{align}
	\bar{h}_{\mu \nu}(t,\boldsymbol{x}) &= \mathcal{H} \Theta'_{\mu \nu} \nonumber \\
	&\times   \biggl\{  t^{2}  + \frac{\cos(2 \eta) - \cos(2\eta')}{8\omega^{2}}  + \frac{t}{2\omega} \sin(2 \eta) \biggr\},
\end{align}
where $\eta' = \omega t + k z$. Noting that $\bar{h}_{\mu}^{\mu} = 0$ as for homogeneous fields, we get $\bar{h}_{\mu \nu} = h_{\mu \nu}$. 

At this point, we use Eq. \eqref{eq:cdt_dz} to evaluate the induced length difference by LIGE. According to Eq. \eqref{eq:h_munu_pw}, we have that $h_{00} = h_{33} = -h_{03} \equiv h$. Reporting this into Eq. \eqref{eq:cdt_dz}, we get
\begin{align}
	c dt &= \frac{(1+ h)}{h + \sqrt{ h^{2} +  (1+h) (1 - h)}} dz, \\
	     &= dz.
\end{align}
Therefore, we conclude that a probe photon co-propagating with a plane-wave will not experience a delay due to LIGE.

\section{Details of the e-dipole model \label{app:time_derivatives}}

The time derivatives of the driving function are given by 
\begin{align}
	\dot{g}(t) &= e^{-(a t)^{2}} \left\{C_{1}(t) \cos(\omega t) + S_{1}(t) \sin(\omega t)\right\} , \\
	\ddot{g}(t) &= e^{-(a t)^{2}} \left\{C_{2}(t) \cos(\omega t) + S_{2}(t) \sin(\omega t)\right\}, \\
	\dddot{g}(t) &= e^{-(a t)^{2}} \left\{ C_{3}(t) \cos(\omega t) + S_{3}(t) \sin(\omega t)\right\} ,
\end{align}
where
\begin{align}
	C_{1}(t) &= \omega , \\
	S_{1}(t) &= - 2 a^{2} t ,\\
	C_{2}(t) &= - 4 a^{2} t\omega ,\\
	S_{2}(t) &=  4a^{4} t^{2} - 2a^{2}  -\omega^{2} ,\\
	C_{3}(t) &= 12a^{4} t^{2} \omega - 6a^{2}\omega  -\omega^{3} , \\
	S_{3}(t) &= 6a^{2}t \omega^{2} - 8 a^{6}t^{3} + 12 a^{4}t .
\end{align}
According to Ref. \cite{PhysRevA.86.053836}, the field at $|\boldsymbol{x}| = 0$, where it reaches its maximum amplitude, is given by
\begin{align}
	\boldsymbol{B}(t,0) &= 0, \\
	\boldsymbol{E}(t,0) &= \frac{1}{4\pi \epsilon_{0}} \frac{4}{3c^{3}} \boldsymbol{d}_{0} \dddot{g}(t).
\end{align}
We have 
\begin{align}
	\label{eq:e_field_max}
	E_{\mathrm{max}} &= |\boldsymbol{E}(0,0)| = \frac{1}{4\pi \epsilon_{0}} \frac{4}{3c^{3}} d_{0} \dddot{g}(0), \\
	&= \frac{1}{4\pi \epsilon_{0}} \frac{4}{3c^{3}} d_{0}  C_{3}(0), \\
	&=  \frac{1}{4\pi \epsilon_{0}} \frac{4 \omega^{3}}{3c^{3}} d_{0} 
	\left[1 + \frac{6a^{2}}{\omega^{2}}    \right].
\end{align}
The energy in the pulse can be evaluated in the far-field $R = R_{0} \rightarrow \infty$. We obtain \cite{PhysRevA.86.053836}
\begin{align}
	U &= \frac{1}{4 \pi \epsilon_{0}} \frac{d^{2}_{0}}{3c^{3}} \int_{-\infty}^{\infty} \ddot{g}^{2}(\tau) d\tau , \\
	&= \frac{1}{4 \pi \epsilon_{0}} \frac{d^{2}_{0}}{3c^{3}} \frac{\omega^{4}}{2a} K(a,\omega)  \\
	K(a,\omega)&= \sqrt{\frac{\pi}{2}}\left[1 + 6\frac{a^{2}}{\omega^{2}} + 3\frac{a^{4}}{\omega^{4}} - 3 \frac{a^{4}}{\omega^{4}} e^{-\frac{\omega^{2}}{2a^{2}}}\right].
\end{align}
Then, the dipole can be written in terms of the pulse energy
\begin{align}
	\label{eq:d0_edip}
	d_{0} &= \sqrt{\frac{24\pi \epsilon_{0} c^{3} a U}{\omega^{4} K(a,\omega)}}.	
\end{align}
This expression of the dipole can be reported in the Eq. \eqref{eq:e_field_max} to get the maximum electric field in terms of pulse energy.


\bibliography{refs}

\begin{thebibliography}{75}%
\makeatletter
\providecommand \@ifxundefined [1]{%
 \@ifx{#1\undefined}
}%
\providecommand \@ifnum [1]{%
 \ifnum #1\expandafter \@firstoftwo
 \else \expandafter \@secondoftwo
 \fi
}%
\providecommand \@ifx [1]{%
 \ifx #1\expandafter \@firstoftwo
 \else \expandafter \@secondoftwo
 \fi
}%
\providecommand \natexlab [1]{#1}%
\providecommand \enquote  [1]{``#1''}%
\providecommand \bibnamefont  [1]{#1}%
\providecommand \bibfnamefont [1]{#1}%
\providecommand \citenamefont [1]{#1}%
\providecommand \href@noop [0]{\@secondoftwo}%
\providecommand \href [0]{\begingroup \@sanitize@url \@href}%
\providecommand \@href[1]{\@@startlink{#1}\@@href}%
\providecommand \@@href[1]{\endgroup#1\@@endlink}%
\providecommand \@sanitize@url [0]{\catcode `\\12\catcode `\$12\catcode
  `\&12\catcode `\#12\catcode `\^12\catcode `\_12\catcode `\%12\relax}%
\providecommand \@@startlink[1]{}%
\providecommand \@@endlink[0]{}%
\providecommand \url  [0]{\begingroup\@sanitize@url \@url }%
\providecommand \@url [1]{\endgroup\@href {#1}{\urlprefix }}%
\providecommand \urlprefix  [0]{URL }%
\providecommand \Eprint [0]{\href }%
\providecommand \doibase [0]{https://doi.org/}%
\providecommand \selectlanguage [0]{\@gobble}%
\providecommand \bibinfo  [0]{\@secondoftwo}%
\providecommand \bibfield  [0]{\@secondoftwo}%
\providecommand \translation [1]{[#1]}%
\providecommand \BibitemOpen [0]{}%
\providecommand \bibitemStop [0]{}%
\providecommand \bibitemNoStop [0]{.\EOS\space}%
\providecommand \EOS [0]{\spacefactor3000\relax}%
\providecommand \BibitemShut  [1]{\csname bibitem#1\endcsname}%
\let\auto@bib@innerbib\@empty
\bibitem [{\citenamefont {Misner}\ \emph {et~al.}(1973)\citenamefont {Misner},
  \citenamefont {Thorne},\ and\ \citenamefont {Wheeler}}]{Misner:1973prb}%
  \BibitemOpen
  \bibfield  {author} {\bibinfo {author} {\bibfnamefont {C.~W.}\ \bibnamefont
  {Misner}}, \bibinfo {author} {\bibfnamefont {K.~S.}\ \bibnamefont {Thorne}},\
  and\ \bibinfo {author} {\bibfnamefont {J.~A.}\ \bibnamefont {Wheeler}},\
  }\href@noop {} {\emph {\bibinfo {title} {{Gravitation}}}}\ (\bibinfo
  {publisher} {W. H. Freeman},\ \bibinfo {address} {San Francisco},\ \bibinfo
  {year} {1973})\BibitemShut {NoStop}%
\bibitem [{\citenamefont {Weinberg}(1972)}]{Weinberg:1972kfs}%
  \BibitemOpen
  \bibfield  {author} {\bibinfo {author} {\bibfnamefont {S.}~\bibnamefont
  {Weinberg}},\ }\href@noop {} {\emph {\bibinfo {title} {{Gravitation and
  Cosmology}: {Principles and Applications of the General Theory of
  Relativity}}}}\ (\bibinfo  {publisher} {John Wiley and Sons},\ \bibinfo
  {address} {New York},\ \bibinfo {year} {1972})\BibitemShut {NoStop}%
\bibitem [{\citenamefont {Carroll}(2019)}]{Carroll:2004st}%
  \BibitemOpen
  \bibfield  {author} {\bibinfo {author} {\bibfnamefont {S.~M.}\ \bibnamefont
  {Carroll}},\ }\href {https://doi.org/10.1017/9781108770385} {\emph {\bibinfo
  {title} {{Spacetime and Geometry}: {An Introduction to General
  Relativity}}}}\ (\bibinfo  {publisher} {Cambridge University Press},\
  \bibinfo {year} {2019})\BibitemShut {NoStop}%
\bibitem [{\citenamefont
  {Einstein}(1916)}]{https://doi.org/10.1002/andp.19163540702}%
  \BibitemOpen
  \bibfield  {author} {\bibinfo {author} {\bibfnamefont {A.}~\bibnamefont
  {Einstein}},\ }\bibfield  {title} {\bibinfo {title} {Die grundlage der
  allgemeinen relativitätstheorie},\ }\href
  {https://doi.org/https://doi.org/10.1002/andp.19163540702} {\bibfield
  {journal} {\bibinfo  {journal} {Annalen der Physik}\ }\textbf {\bibinfo
  {volume} {354}},\ \bibinfo {pages} {769} (\bibinfo {year}
  {1916})}\BibitemShut {NoStop}%
\bibitem [{\citenamefont {Stewart}(2005)}]{10.1119/1.1949625}%
  \BibitemOpen
  \bibfield  {author} {\bibinfo {author} {\bibfnamefont {M.~G.}\ \bibnamefont
  {Stewart}},\ }\bibfield  {title} {\bibinfo {title} {{Precession of the
  perihelion of Mercury’s orbit}},\ }\href
  {https://doi.org/10.1119/1.1949625} {\bibfield  {journal} {\bibinfo
  {journal} {American Journal of Physics}\ }\textbf {\bibinfo {volume} {73}},\
  \bibinfo {pages} {730} (\bibinfo {year} {2005})}\BibitemShut {NoStop}%
\bibitem [{\citenamefont {Park}\ \emph {et~al.}(2017)\citenamefont {Park},
  \citenamefont {Folkner}, \citenamefont {Konopliv}, \citenamefont {Williams},
  \citenamefont {Smith},\ and\ \citenamefont {Zuber}}]{Park_2017}%
  \BibitemOpen
  \bibfield  {author} {\bibinfo {author} {\bibfnamefont {R.~S.}\ \bibnamefont
  {Park}}, \bibinfo {author} {\bibfnamefont {W.~M.}\ \bibnamefont {Folkner}},
  \bibinfo {author} {\bibfnamefont {A.~S.}\ \bibnamefont {Konopliv}}, \bibinfo
  {author} {\bibfnamefont {J.~G.}\ \bibnamefont {Williams}}, \bibinfo {author}
  {\bibfnamefont {D.~E.}\ \bibnamefont {Smith}},\ and\ \bibinfo {author}
  {\bibfnamefont {M.~T.}\ \bibnamefont {Zuber}},\ }\bibfield  {title} {\bibinfo
  {title} {Precession of mercury’s perihelion from ranging to the messenger
  spacecraft},\ }\href {https://doi.org/10.3847/1538-3881/aa5be2} {\bibfield
  {journal} {\bibinfo  {journal} {The Astronomical Journal}\ }\textbf {\bibinfo
  {volume} {153}},\ \bibinfo {pages} {121} (\bibinfo {year}
  {2017})}\BibitemShut {NoStop}%
\bibitem [{\citenamefont {Shapiro}\ \emph {et~al.}(2004)\citenamefont
  {Shapiro}, \citenamefont {Davis}, \citenamefont {Lebach},\ and\ \citenamefont
  {Gregory}}]{PhysRevLett.92.121101}%
  \BibitemOpen
  \bibfield  {author} {\bibinfo {author} {\bibfnamefont {S.~S.}\ \bibnamefont
  {Shapiro}}, \bibinfo {author} {\bibfnamefont {J.~L.}\ \bibnamefont {Davis}},
  \bibinfo {author} {\bibfnamefont {D.~E.}\ \bibnamefont {Lebach}},\ and\
  \bibinfo {author} {\bibfnamefont {J.~S.}\ \bibnamefont {Gregory}},\
  }\bibfield  {title} {\bibinfo {title} {Measurement of the solar gravitational
  deflection of radio waves using geodetic very-long-baseline interferometry
  data, 1979--1999},\ }\href {https://doi.org/10.1103/PhysRevLett.92.121101}
  {\bibfield  {journal} {\bibinfo  {journal} {Phys. Rev. Lett.}\ }\textbf
  {\bibinfo {volume} {92}},\ \bibinfo {pages} {121101} (\bibinfo {year}
  {2004})}\BibitemShut {NoStop}%
\bibitem [{\citenamefont {Fomalont}\ \emph {et~al.}(2009)\citenamefont
  {Fomalont}, \citenamefont {Kopeikin}, \citenamefont {Lanyi},\ and\
  \citenamefont {Benson}}]{Fomalont_2009}%
  \BibitemOpen
  \bibfield  {author} {\bibinfo {author} {\bibfnamefont {E.}~\bibnamefont
  {Fomalont}}, \bibinfo {author} {\bibfnamefont {S.}~\bibnamefont {Kopeikin}},
  \bibinfo {author} {\bibfnamefont {G.}~\bibnamefont {Lanyi}},\ and\ \bibinfo
  {author} {\bibfnamefont {J.}~\bibnamefont {Benson}},\ }\bibfield  {title}
  {\bibinfo {title} {Progress in measurements of the gravitational bending of
  radio waves using the vlba},\ }\href
  {https://doi.org/10.1088/0004-637X/699/2/1395} {\bibfield  {journal}
  {\bibinfo  {journal} {The Astrophysical Journal}\ }\textbf {\bibinfo {volume}
  {699}},\ \bibinfo {pages} {1395} (\bibinfo {year} {2009})}\BibitemShut
  {NoStop}%
\bibitem [{\citenamefont {Earman}\ and\ \citenamefont
  {Glymour}(1980)}]{EARMAN1980175}%
  \BibitemOpen
  \bibfield  {author} {\bibinfo {author} {\bibfnamefont {J.}~\bibnamefont
  {Earman}}\ and\ \bibinfo {author} {\bibfnamefont {C.}~\bibnamefont
  {Glymour}},\ }\bibfield  {title} {\bibinfo {title} {The gravitational red
  shift as a test of general relativity: History and analysis},\ }\href
  {https://doi.org/https://doi.org/10.1016/0039-3681(80)90025-4} {\bibfield
  {journal} {\bibinfo  {journal} {Studies in History and Philosophy of Science
  Part A}\ }\textbf {\bibinfo {volume} {11}},\ \bibinfo {pages} {175} (\bibinfo
  {year} {1980})}\BibitemShut {NoStop}%
\bibitem [{\citenamefont {Vessot}\ \emph {et~al.}(1980)\citenamefont {Vessot},
  \citenamefont {Levine}, \citenamefont {Mattison}, \citenamefont {Blomberg},
  \citenamefont {Hoffman}, \citenamefont {Nystrom}, \citenamefont {Farrel},
  \citenamefont {Decher}, \citenamefont {Eby}, \citenamefont {Baugher},
  \citenamefont {Watts}, \citenamefont {Teuber},\ and\ \citenamefont
  {Wills}}]{PhysRevLett.45.2081}%
  \BibitemOpen
  \bibfield  {author} {\bibinfo {author} {\bibfnamefont {R.~F.~C.}\
  \bibnamefont {Vessot}}, \bibinfo {author} {\bibfnamefont {M.~W.}\
  \bibnamefont {Levine}}, \bibinfo {author} {\bibfnamefont {E.~M.}\
  \bibnamefont {Mattison}}, \bibinfo {author} {\bibfnamefont {E.~L.}\
  \bibnamefont {Blomberg}}, \bibinfo {author} {\bibfnamefont {T.~E.}\
  \bibnamefont {Hoffman}}, \bibinfo {author} {\bibfnamefont {G.~U.}\
  \bibnamefont {Nystrom}}, \bibinfo {author} {\bibfnamefont {B.~F.}\
  \bibnamefont {Farrel}}, \bibinfo {author} {\bibfnamefont {R.}~\bibnamefont
  {Decher}}, \bibinfo {author} {\bibfnamefont {P.~B.}\ \bibnamefont {Eby}},
  \bibinfo {author} {\bibfnamefont {C.~R.}\ \bibnamefont {Baugher}}, \bibinfo
  {author} {\bibfnamefont {J.~W.}\ \bibnamefont {Watts}}, \bibinfo {author}
  {\bibfnamefont {D.~L.}\ \bibnamefont {Teuber}},\ and\ \bibinfo {author}
  {\bibfnamefont {F.~D.}\ \bibnamefont {Wills}},\ }\bibfield  {title} {\bibinfo
  {title} {Test of relativistic gravitation with a space-borne hydrogen
  maser},\ }\href {https://doi.org/10.1103/PhysRevLett.45.2081} {\bibfield
  {journal} {\bibinfo  {journal} {Phys. Rev. Lett.}\ }\textbf {\bibinfo
  {volume} {45}},\ \bibinfo {pages} {2081} (\bibinfo {year}
  {1980})}\BibitemShut {NoStop}%
\bibitem [{\citenamefont {M{\"u}ller}\ \emph {et~al.}(2010)\citenamefont
  {M{\"u}ller}, \citenamefont {Peters},\ and\ \citenamefont
  {Chu}}]{muller2010precision}%
  \BibitemOpen
  \bibfield  {author} {\bibinfo {author} {\bibfnamefont {H.}~\bibnamefont
  {M{\"u}ller}}, \bibinfo {author} {\bibfnamefont {A.}~\bibnamefont {Peters}},\
  and\ \bibinfo {author} {\bibfnamefont {S.}~\bibnamefont {Chu}},\ }\bibfield
  {title} {\bibinfo {title} {A precision measurement of the gravitational
  redshift by the interference of matter waves},\ }\href
  {https://doi.org/https://doi.org/10.1038/nature08776} {\bibfield  {journal}
  {\bibinfo  {journal} {Nature}\ }\textbf {\bibinfo {volume} {463}},\ \bibinfo
  {pages} {926} (\bibinfo {year} {2010})}\BibitemShut {NoStop}%
\bibitem [{\citenamefont {Akiyama}\ \emph {et~al.}(2019)\citenamefont {Akiyama}
  \emph {et~al.}}]{Akiyama_2019}%
  \BibitemOpen
  \bibfield  {author} {\bibinfo {author} {\bibfnamefont {K.}~\bibnamefont
  {Akiyama}} \emph {et~al.} (\bibinfo {collaboration} {The Event Horizon
  Telescope Collaboration}),\ }\bibfield  {title} {\bibinfo {title} {First m87
  event horizon telescope results. i. the shadow of the supermassive black
  hole},\ }\href {https://doi.org/10.3847/2041-8213/ab0ec7} {\bibfield
  {journal} {\bibinfo  {journal} {The Astrophysical Journal Letters}\ }\textbf
  {\bibinfo {volume} {875}},\ \bibinfo {pages} {L1} (\bibinfo {year}
  {2019})}\BibitemShut {NoStop}%
\bibitem [{\citenamefont {Maggiore}(2007)}]{maggiore_book}%
  \BibitemOpen
  \bibfield  {author} {\bibinfo {author} {\bibfnamefont {M.}~\bibnamefont
  {Maggiore}},\ }\href
  {https://doi.org/10.1093/acprof:oso/9780198570745.001.0001} {\emph {\bibinfo
  {title} {{Gravitational Waves: Volume 1: Theory and Experiments}}}}\
  (\bibinfo  {publisher} {Oxford University Press},\ \bibinfo {year}
  {2007})\BibitemShut {NoStop}%
\bibitem [{\citenamefont {Abbott}\ \emph
  {et~al.}(2016{\natexlab{a}})\citenamefont {Abbott} \emph
  {et~al.}}]{PhysRevLett.116.061102}%
  \BibitemOpen
  \bibfield  {author} {\bibinfo {author} {\bibfnamefont {B.~P.}\ \bibnamefont
  {Abbott}} \emph {et~al.} (\bibinfo {collaboration} {LIGO Scientific
  Collaboration and Virgo Collaboration}),\ }\bibfield  {title} {\bibinfo
  {title} {Observation of gravitational waves from a binary black hole
  merger},\ }\href {https://doi.org/10.1103/PhysRevLett.116.061102} {\bibfield
  {journal} {\bibinfo  {journal} {Phys. Rev. Lett.}\ }\textbf {\bibinfo
  {volume} {116}},\ \bibinfo {pages} {061102} (\bibinfo {year}
  {2016}{\natexlab{a}})}\BibitemShut {NoStop}%
\bibitem [{\citenamefont {Abbott}\ \emph
  {et~al.}(2016{\natexlab{b}})\citenamefont {Abbott} \emph
  {et~al.}}]{PhysRevLett.116.241103}%
  \BibitemOpen
  \bibfield  {author} {\bibinfo {author} {\bibfnamefont {B.~P.}\ \bibnamefont
  {Abbott}} \emph {et~al.} (\bibinfo {collaboration} {LIGO Scientific
  Collaboration and Virgo Collaboration}),\ }\bibfield  {title} {\bibinfo
  {title} {Gw151226: Observation of gravitational waves from a 22-solar-mass
  binary black hole coalescence},\ }\href
  {https://doi.org/10.1103/PhysRevLett.116.241103} {\bibfield  {journal}
  {\bibinfo  {journal} {Phys. Rev. Lett.}\ }\textbf {\bibinfo {volume} {116}},\
  \bibinfo {pages} {241103} (\bibinfo {year} {2016}{\natexlab{b}})}\BibitemShut
  {NoStop}%
\bibitem [{\citenamefont {Wheeler}(1955)}]{PhysRev.97.511}%
  \BibitemOpen
  \bibfield  {author} {\bibinfo {author} {\bibfnamefont {J.~A.}\ \bibnamefont
  {Wheeler}},\ }\bibfield  {title} {\bibinfo {title} {Geons},\ }\href
  {https://doi.org/10.1103/PhysRev.97.511} {\bibfield  {journal} {\bibinfo
  {journal} {Phys. Rev.}\ }\textbf {\bibinfo {volume} {97}},\ \bibinfo {pages}
  {511} (\bibinfo {year} {1955})}\BibitemShut {NoStop}%
\bibitem [{\citenamefont {Ernst}(1957)}]{PhysRev.105.1665}%
  \BibitemOpen
  \bibfield  {author} {\bibinfo {author} {\bibfnamefont {F.~J.}\ \bibnamefont
  {Ernst}},\ }\bibfield  {title} {\bibinfo {title} {Linear and toroidal
  geons},\ }\href {https://doi.org/10.1103/PhysRev.105.1665} {\bibfield
  {journal} {\bibinfo  {journal} {Phys. Rev.}\ }\textbf {\bibinfo {volume}
  {105}},\ \bibinfo {pages} {1665} (\bibinfo {year} {1957})}\BibitemShut
  {NoStop}%
\bibitem [{\citenamefont {Senovilla}(2014)}]{Senovilla_2015}%
  \BibitemOpen
  \bibfield  {author} {\bibinfo {author} {\bibfnamefont {J.~M.~M.}\
  \bibnamefont {Senovilla}},\ }\bibfield  {title} {\bibinfo {title} {Black hole
  formation by incoming electromagnetic radiation},\ }\href
  {https://doi.org/10.1088/0264-9381/32/1/017001} {\bibfield  {journal}
  {\bibinfo  {journal} {Classical and Quantum Gravity}\ }\textbf {\bibinfo
  {volume} {32}},\ \bibinfo {pages} {017001} (\bibinfo {year}
  {2014})}\BibitemShut {NoStop}%
\bibitem [{\citenamefont {\'Alvarez-Dom\'{\i}nguez}\ \emph
  {et~al.}(2024)\citenamefont {\'Alvarez-Dom\'{\i}nguez}, \citenamefont
  {Garay}, \citenamefont {Mart\'{\i}n-Mart\'{\i}nez},\ and\ \citenamefont
  {Polo-G\'omez}}]{PhysRevLett.133.041401}%
  \BibitemOpen
  \bibfield  {author} {\bibinfo {author} {\bibfnamefont {A.}~\bibnamefont
  {\'Alvarez-Dom\'{\i}nguez}}, \bibinfo {author} {\bibfnamefont {L.~J.}\
  \bibnamefont {Garay}}, \bibinfo {author} {\bibfnamefont {E.}~\bibnamefont
  {Mart\'{\i}n-Mart\'{\i}nez}},\ and\ \bibinfo {author} {\bibfnamefont
  {J.}~\bibnamefont {Polo-G\'omez}},\ }\bibfield  {title} {\bibinfo {title} {No
  black holes from light},\ }\href
  {https://doi.org/10.1103/PhysRevLett.133.041401} {\bibfield  {journal}
  {\bibinfo  {journal} {Phys. Rev. Lett.}\ }\textbf {\bibinfo {volume} {133}},\
  \bibinfo {pages} {041401} (\bibinfo {year} {2024})}\BibitemShut {NoStop}%
\bibitem [{\citenamefont {Gertsenshtein}(1962)}]{gertsenshtein1962wave}%
  \BibitemOpen
  \bibfield  {author} {\bibinfo {author} {\bibfnamefont {M.}~\bibnamefont
  {Gertsenshtein}},\ }\bibfield  {title} {\bibinfo {title} {Wave resonance of
  light and gravitional waves},\ }\href@noop {} {\bibfield  {journal} {\bibinfo
   {journal} {Sov Phys JETP}\ }\textbf {\bibinfo {volume} {14}},\ \bibinfo
  {pages} {84} (\bibinfo {year} {1962})}\BibitemShut {NoStop}%
\bibitem [{\citenamefont {De~Logi}\ and\ \citenamefont
  {Mickelson}(1977)}]{PhysRevD.16.2915}%
  \BibitemOpen
  \bibfield  {author} {\bibinfo {author} {\bibfnamefont {W.~K.}\ \bibnamefont
  {De~Logi}}\ and\ \bibinfo {author} {\bibfnamefont {A.~R.}\ \bibnamefont
  {Mickelson}},\ }\bibfield  {title} {\bibinfo {title} {Electrogravitational
  conversion cross sections in static electromagnetic fields},\ }\href
  {https://doi.org/10.1103/PhysRevD.16.2915} {\bibfield  {journal} {\bibinfo
  {journal} {Phys. Rev. D}\ }\textbf {\bibinfo {volume} {16}},\ \bibinfo
  {pages} {2915} (\bibinfo {year} {1977})}\BibitemShut {NoStop}%
\bibitem [{\citenamefont {Long}\ \emph {et~al.}(1994)\citenamefont {Long},
  \citenamefont {Soa},\ and\ \citenamefont
  {Tran}}]{doi:10.1142/S0217732394003464}%
  \BibitemOpen
  \bibfield  {author} {\bibinfo {author} {\bibfnamefont {H.~N.}\ \bibnamefont
  {Long}}, \bibinfo {author} {\bibfnamefont {D.~V.}\ \bibnamefont {Soa}},\ and\
  \bibinfo {author} {\bibfnamefont {T.~A.}\ \bibnamefont {Tran}},\ }\bibfield
  {title} {\bibinfo {title} {Electromagnetic-gravitational conversion
  cross-sections in external electromagnetic fields},\ }\href
  {https://doi.org/10.1142/S0217732394003464} {\bibfield  {journal} {\bibinfo
  {journal} {Modern Physics Letters A}\ }\textbf {\bibinfo {volume} {09}},\
  \bibinfo {pages} {3619} (\bibinfo {year} {1994})}\BibitemShut {NoStop}%
\bibitem [{\citenamefont {Palessandro}\ and\ \citenamefont
  {Rothman}(2023)}]{PALESSANDRO2023101187}%
  \BibitemOpen
  \bibfield  {author} {\bibinfo {author} {\bibfnamefont {A.}~\bibnamefont
  {Palessandro}}\ and\ \bibinfo {author} {\bibfnamefont {T.}~\bibnamefont
  {Rothman}},\ }\bibfield  {title} {\bibinfo {title} {A simple derivation of
  the gertsenshtein effect},\ }\href
  {https://doi.org/https://doi.org/10.1016/j.dark.2023.101187} {\bibfield
  {journal} {\bibinfo  {journal} {Physics of the Dark Universe}\ }\textbf
  {\bibinfo {volume} {40}},\ \bibinfo {pages} {101187} (\bibinfo {year}
  {2023})}\BibitemShut {NoStop}%
\bibitem [{\citenamefont {Raffelt}\ and\ \citenamefont
  {Stodolsky}(1988)}]{PhysRevD.37.1237}%
  \BibitemOpen
  \bibfield  {author} {\bibinfo {author} {\bibfnamefont {G.}~\bibnamefont
  {Raffelt}}\ and\ \bibinfo {author} {\bibfnamefont {L.}~\bibnamefont
  {Stodolsky}},\ }\bibfield  {title} {\bibinfo {title} {Mixing of the photon
  with low-mass particles},\ }\href {https://doi.org/10.1103/PhysRevD.37.1237}
  {\bibfield  {journal} {\bibinfo  {journal} {Phys. Rev. D}\ }\textbf {\bibinfo
  {volume} {37}},\ \bibinfo {pages} {1237} (\bibinfo {year}
  {1988})}\BibitemShut {NoStop}%
\bibitem [{\citenamefont {Magueijo}(1994)}]{PhysRevD.49.671}%
  \BibitemOpen
  \bibfield  {author} {\bibinfo {author} {\bibfnamefont {J.~C.~R.}\
  \bibnamefont {Magueijo}},\ }\bibfield  {title} {\bibinfo {title} {Cosmic
  magnetic field imprints on cosmic radiation},\ }\href
  {https://doi.org/10.1103/PhysRevD.49.671} {\bibfield  {journal} {\bibinfo
  {journal} {Phys. Rev. D}\ }\textbf {\bibinfo {volume} {49}},\ \bibinfo
  {pages} {671} (\bibinfo {year} {1994})}\BibitemShut {NoStop}%
\bibitem [{\citenamefont {Braginsky}\ \emph {et~al.}(1977)\citenamefont
  {Braginsky}, \citenamefont {Caves},\ and\ \citenamefont
  {Thorne}}]{PhysRevD.15.2047}%
  \BibitemOpen
  \bibfield  {author} {\bibinfo {author} {\bibfnamefont {V.~B.}\ \bibnamefont
  {Braginsky}}, \bibinfo {author} {\bibfnamefont {C.~M.}\ \bibnamefont
  {Caves}},\ and\ \bibinfo {author} {\bibfnamefont {K.~S.}\ \bibnamefont
  {Thorne}},\ }\bibfield  {title} {\bibinfo {title} {Laboratory experiments to
  test relativistic gravity},\ }\href
  {https://doi.org/10.1103/PhysRevD.15.2047} {\bibfield  {journal} {\bibinfo
  {journal} {Phys. Rev. D}\ }\textbf {\bibinfo {volume} {15}},\ \bibinfo
  {pages} {2047} (\bibinfo {year} {1977})}\BibitemShut {NoStop}%
\bibitem [{\citenamefont {Bailey}(2024)}]{Bailey2024}%
  \BibitemOpen
  \bibfield  {author} {\bibinfo {author} {\bibfnamefont {Q.~G.}\ \bibnamefont
  {Bailey}},\ }\bibinfo {title} {Testing gravity in the laboratory},\ in\ \href
  {https://doi.org/10.1007/978-981-97-2871-8_1} {\emph {\bibinfo {booktitle}
  {Recent Progress on Gravity Tests: Challenges and Future Perspectives}}},\
  \bibinfo {editor} {edited by\ \bibinfo {editor} {\bibfnamefont
  {C.}~\bibnamefont {Bambi}}\ and\ \bibinfo {editor} {\bibfnamefont
  {A.}~\bibnamefont {C{\'a}rdenas-Avenda{\~{n}}o}}}\ (\bibinfo  {publisher}
  {Springer Nature Singapore},\ \bibinfo {address} {Singapore},\ \bibinfo
  {year} {2024})\ pp.\ \bibinfo {pages} {1--26}\BibitemShut {NoStop}%
\bibitem [{\citenamefont {Bahk}\ \emph {et~al.}(2004)\citenamefont {Bahk},
  \citenamefont {Rousseau}, \citenamefont {Planchon}, \citenamefont {Chvykov},
  \citenamefont {Kalintchenko}, \citenamefont {Maksimchuk}, \citenamefont
  {Mourou},\ and\ \citenamefont {Yanovsky}}]{Bahk:04}%
  \BibitemOpen
  \bibfield  {author} {\bibinfo {author} {\bibfnamefont {S.-W.}\ \bibnamefont
  {Bahk}}, \bibinfo {author} {\bibfnamefont {P.}~\bibnamefont {Rousseau}},
  \bibinfo {author} {\bibfnamefont {T.~A.}\ \bibnamefont {Planchon}}, \bibinfo
  {author} {\bibfnamefont {V.}~\bibnamefont {Chvykov}}, \bibinfo {author}
  {\bibfnamefont {G.}~\bibnamefont {Kalintchenko}}, \bibinfo {author}
  {\bibfnamefont {A.}~\bibnamefont {Maksimchuk}}, \bibinfo {author}
  {\bibfnamefont {G.~A.}\ \bibnamefont {Mourou}},\ and\ \bibinfo {author}
  {\bibfnamefont {V.}~\bibnamefont {Yanovsky}},\ }\bibfield  {title} {\bibinfo
  {title} {Generation and characterization of the highest laser intensities
  (10$^{22}$ w/cm$^2$)},\ }\href {https://doi.org/10.1364/OL.29.002837}
  {\bibfield  {journal} {\bibinfo  {journal} {Opt. Lett.}\ }\textbf {\bibinfo
  {volume} {29}},\ \bibinfo {pages} {2837} (\bibinfo {year}
  {2004})}\BibitemShut {NoStop}%
\bibitem [{\citenamefont {Yoon}\ \emph {et~al.}(2021)\citenamefont {Yoon},
  \citenamefont {Kim}, \citenamefont {Choi}, \citenamefont {Sung},
  \citenamefont {Lee}, \citenamefont {Lee},\ and\ \citenamefont
  {Nam}}]{Yoon:21}%
  \BibitemOpen
  \bibfield  {author} {\bibinfo {author} {\bibfnamefont {J.~W.}\ \bibnamefont
  {Yoon}}, \bibinfo {author} {\bibfnamefont {Y.~G.}\ \bibnamefont {Kim}},
  \bibinfo {author} {\bibfnamefont {I.~W.}\ \bibnamefont {Choi}}, \bibinfo
  {author} {\bibfnamefont {J.~H.}\ \bibnamefont {Sung}}, \bibinfo {author}
  {\bibfnamefont {H.~W.}\ \bibnamefont {Lee}}, \bibinfo {author} {\bibfnamefont
  {S.~K.}\ \bibnamefont {Lee}},\ and\ \bibinfo {author} {\bibfnamefont {C.~H.}\
  \bibnamefont {Nam}},\ }\bibfield  {title} {\bibinfo {title} {Realization of
  laser intensity over 10$^{23}$ w/cm$^2$},\ }\href
  {https://doi.org/10.1364/OPTICA.420520} {\bibfield  {journal} {\bibinfo
  {journal} {Optica}\ }\textbf {\bibinfo {volume} {8}},\ \bibinfo {pages} {630}
  (\bibinfo {year} {2021})}\BibitemShut {NoStop}%
\bibitem [{\citenamefont {Danson}\ \emph {et~al.}(2019)\citenamefont {Danson},
  \citenamefont {Haefner}, \citenamefont {Bromage}, \citenamefont {Butcher},
  \citenamefont {Chanteloup}, \citenamefont {Chowdhury}, \citenamefont
  {Galvanauskas}, \citenamefont {Gizzi}, \citenamefont {Hein}, \citenamefont
  {Hillier},\ and\ \citenamefont {et~al.}}]{Danson_al_2019}%
  \BibitemOpen
  \bibfield  {author} {\bibinfo {author} {\bibfnamefont {C.~N.}\ \bibnamefont
  {Danson}}, \bibinfo {author} {\bibfnamefont {C.}~\bibnamefont {Haefner}},
  \bibinfo {author} {\bibfnamefont {J.}~\bibnamefont {Bromage}}, \bibinfo
  {author} {\bibfnamefont {T.}~\bibnamefont {Butcher}}, \bibinfo {author}
  {\bibfnamefont {J.-C.~F.}\ \bibnamefont {Chanteloup}}, \bibinfo {author}
  {\bibfnamefont {E.~A.}\ \bibnamefont {Chowdhury}}, \bibinfo {author}
  {\bibfnamefont {A.}~\bibnamefont {Galvanauskas}}, \bibinfo {author}
  {\bibfnamefont {L.~A.}\ \bibnamefont {Gizzi}}, \bibinfo {author}
  {\bibfnamefont {J.}~\bibnamefont {Hein}}, \bibinfo {author} {\bibfnamefont
  {D.~I.}\ \bibnamefont {Hillier}},\ and\ \bibinfo {author} {\bibnamefont
  {et~al.}},\ }\bibfield  {title} {\bibinfo {title} {Petawatt and exawatt class
  lasers worldwide},\ }\href {https://doi.org/10.1017/hpl.2019.36} {\bibfield
  {journal} {\bibinfo  {journal} {High Power Laser Science and Engineering}\
  }\textbf {\bibinfo {volume} {7}},\ \bibinfo {pages} {e54} (\bibinfo {year}
  {2019})}\BibitemShut {NoStop}%
\bibitem [{\citenamefont {Tolman}\ \emph {et~al.}(1931)\citenamefont {Tolman},
  \citenamefont {Ehrenfest},\ and\ \citenamefont {Podolsky}}]{PhysRev.37.602}%
  \BibitemOpen
  \bibfield  {author} {\bibinfo {author} {\bibfnamefont {R.~C.}\ \bibnamefont
  {Tolman}}, \bibinfo {author} {\bibfnamefont {P.}~\bibnamefont {Ehrenfest}},\
  and\ \bibinfo {author} {\bibfnamefont {B.}~\bibnamefont {Podolsky}},\
  }\bibfield  {title} {\bibinfo {title} {On the gravitational field produced by
  light},\ }\href {https://doi.org/10.1103/PhysRev.37.602} {\bibfield
  {journal} {\bibinfo  {journal} {Phys. Rev.}\ }\textbf {\bibinfo {volume}
  {37}},\ \bibinfo {pages} {602} (\bibinfo {year} {1931})}\BibitemShut
  {NoStop}%
\bibitem [{\citenamefont {Adler}(1975)}]{PhysRevD.11.2685}%
  \BibitemOpen
  \bibfield  {author} {\bibinfo {author} {\bibfnamefont {R.~J.}\ \bibnamefont
  {Adler}},\ }\bibfield  {title} {\bibinfo {title} {Gravitational radiation
  from laser pulses},\ }\href {https://doi.org/10.1103/PhysRevD.11.2685}
  {\bibfield  {journal} {\bibinfo  {journal} {Phys. Rev. D}\ }\textbf {\bibinfo
  {volume} {11}},\ \bibinfo {pages} {2685} (\bibinfo {year}
  {1975})}\BibitemShut {NoStop}%
\bibitem [{\citenamefont {Scully}(1979)}]{PhysRevD.19.3582}%
  \BibitemOpen
  \bibfield  {author} {\bibinfo {author} {\bibfnamefont {M.~O.}\ \bibnamefont
  {Scully}},\ }\bibfield  {title} {\bibinfo {title} {General-relativistic
  treatment of the gravitational coupling between laser beams},\ }\href
  {https://doi.org/10.1103/PhysRevD.19.3582} {\bibfield  {journal} {\bibinfo
  {journal} {Phys. Rev. D}\ }\textbf {\bibinfo {volume} {19}},\ \bibinfo
  {pages} {3582} (\bibinfo {year} {1979})}\BibitemShut {NoStop}%
\bibitem [{\citenamefont {Mallett}(2000)}]{MALLETT2000214}%
  \BibitemOpen
  \bibfield  {author} {\bibinfo {author} {\bibfnamefont {R.~L.}\ \bibnamefont
  {Mallett}},\ }\bibfield  {title} {\bibinfo {title} {Weak gravitational field
  of the electromagnetic radiation in a ring laser},\ }\href
  {https://doi.org/https://doi.org/10.1016/S0375-9601(00)00260-7} {\bibfield
  {journal} {\bibinfo  {journal} {Physics Letters A}\ }\textbf {\bibinfo
  {volume} {269}},\ \bibinfo {pages} {214} (\bibinfo {year}
  {2000})}\BibitemShut {NoStop}%
\bibitem [{\citenamefont {Baker~Jr.}\ \emph {et~al.}(2006)\citenamefont
  {Baker~Jr.}, \citenamefont {Li},\ and\ \citenamefont
  {Li}}]{10.1063/1.2169320}%
  \BibitemOpen
  \bibfield  {author} {\bibinfo {author} {\bibfnamefont {R.~M.~L.}\
  \bibnamefont {Baker~Jr.}}, \bibinfo {author} {\bibfnamefont {F.}~\bibnamefont
  {Li}},\ and\ \bibinfo {author} {\bibfnamefont {R.}~\bibnamefont {Li}},\
  }\bibfield  {title} {\bibinfo {title} {{Ultra‐High‐Intensity Lasers for
  Gravitational Wave Generation and Detection}},\ }\href
  {https://doi.org/10.1063/1.2169320} {\bibfield  {journal} {\bibinfo
  {journal} {AIP Conference Proceedings}\ }\textbf {\bibinfo {volume} {813}},\
  \bibinfo {pages} {1352} (\bibinfo {year} {2006})}\BibitemShut {NoStop}%
\bibitem [{\citenamefont {Ji}\ and\ \citenamefont
  {Bai}(2006)}]{ji2006gravitational}%
  \BibitemOpen
  \bibfield  {author} {\bibinfo {author} {\bibfnamefont {P.}~\bibnamefont
  {Ji}}\ and\ \bibinfo {author} {\bibfnamefont {Y.}~\bibnamefont {Bai}},\
  }\bibfield  {title} {\bibinfo {title} {Gravitational effects induced by
  high-power lasers},\ }\href
  {https://doi.org/https://doi.org/10.1140/epjc/s2006-02531-9} {\bibfield
  {journal} {\bibinfo  {journal} {The European Physical Journal C-Particles and
  Fields}\ }\textbf {\bibinfo {volume} {46}},\ \bibinfo {pages} {817} (\bibinfo
  {year} {2006})}\BibitemShut {NoStop}%
\bibitem [{\citenamefont {Rätzel}\ \emph {et~al.}(2016)\citenamefont
  {Rätzel}, \citenamefont {Wilkens},\ and\ \citenamefont
  {Menzel}}]{Ratzel_2016}%
  \BibitemOpen
  \bibfield  {author} {\bibinfo {author} {\bibfnamefont {D.}~\bibnamefont
  {Rätzel}}, \bibinfo {author} {\bibfnamefont {M.}~\bibnamefont {Wilkens}},\
  and\ \bibinfo {author} {\bibfnamefont {R.}~\bibnamefont {Menzel}},\
  }\bibfield  {title} {\bibinfo {title} {Gravitational properties of
  light—the gravitational field of a laser pulse},\ }\href
  {https://doi.org/10.1088/1367-2630/18/2/023009} {\bibfield  {journal}
  {\bibinfo  {journal} {New Journal of Physics}\ }\textbf {\bibinfo {volume}
  {18}},\ \bibinfo {pages} {023009} (\bibinfo {year} {2016})}\BibitemShut
  {NoStop}%
\bibitem [{\citenamefont {Schneiter}\ \emph {et~al.}(2018)\citenamefont
  {Schneiter}, \citenamefont {Rätzel},\ and\ \citenamefont
  {Braun}}]{Schneiter_2018}%
  \BibitemOpen
  \bibfield  {author} {\bibinfo {author} {\bibfnamefont {F.}~\bibnamefont
  {Schneiter}}, \bibinfo {author} {\bibfnamefont {D.}~\bibnamefont {Rätzel}},\
  and\ \bibinfo {author} {\bibfnamefont {D.}~\bibnamefont {Braun}},\ }\bibfield
   {title} {\bibinfo {title} {The gravitational field of a laser beam beyond
  the short wavelength approximation},\ }\href
  {https://doi.org/10.1088/1361-6382/aadc81} {\bibfield  {journal} {\bibinfo
  {journal} {Classical and Quantum Gravity}\ }\textbf {\bibinfo {volume}
  {35}},\ \bibinfo {pages} {195007} (\bibinfo {year} {2018})}\BibitemShut
  {NoStop}%
\bibitem [{\citenamefont {Lageyre}\ \emph {et~al.}(2022)\citenamefont
  {Lageyre}, \citenamefont {d'Humi\`eres},\ and\ \citenamefont
  {Ribeyre}}]{PhysRevD.105.104052}%
  \BibitemOpen
  \bibfield  {author} {\bibinfo {author} {\bibfnamefont {P.}~\bibnamefont
  {Lageyre}}, \bibinfo {author} {\bibfnamefont {E.}~\bibnamefont
  {d'Humi\`eres}},\ and\ \bibinfo {author} {\bibfnamefont {X.}~\bibnamefont
  {Ribeyre}},\ }\bibfield  {title} {\bibinfo {title} {Gravitational influence
  of high power laser pulses},\ }\href
  {https://doi.org/10.1103/PhysRevD.105.104052} {\bibfield  {journal} {\bibinfo
   {journal} {Phys. Rev. D}\ }\textbf {\bibinfo {volume} {105}},\ \bibinfo
  {pages} {104052} (\bibinfo {year} {2022})}\BibitemShut {NoStop}%
\bibitem [{\citenamefont {Morozov}\ \emph {et~al.}(2021)\citenamefont
  {Morozov}, \citenamefont {Pustovoit},\ and\ \citenamefont
  {Fomin}}]{morozov2021generation}%
  \BibitemOpen
  \bibfield  {author} {\bibinfo {author} {\bibfnamefont {A.}~\bibnamefont
  {Morozov}}, \bibinfo {author} {\bibfnamefont {V.}~\bibnamefont {Pustovoit}},\
  and\ \bibinfo {author} {\bibfnamefont {I.}~\bibnamefont {Fomin}},\ }\bibfield
   {title} {\bibinfo {title} {Generation of gravitational waves by a standing
  electromagnetic wave},\ }\href
  {https://doi.org/https://doi.org/10.1134/S020228932101014X} {\bibfield
  {journal} {\bibinfo  {journal} {Gravitation and Cosmology}\ }\textbf
  {\bibinfo {volume} {27}},\ \bibinfo {pages} {24} (\bibinfo {year}
  {2021})}\BibitemShut {NoStop}%
\bibitem [{\citenamefont {Spengler}\ \emph {et~al.}(2022)\citenamefont
  {Spengler}, \citenamefont {Rätzel},\ and\ \citenamefont
  {Braun}}]{Spengler_2022}%
  \BibitemOpen
  \bibfield  {author} {\bibinfo {author} {\bibfnamefont {F.}~\bibnamefont
  {Spengler}}, \bibinfo {author} {\bibfnamefont {D.}~\bibnamefont {Rätzel}},\
  and\ \bibinfo {author} {\bibfnamefont {D.}~\bibnamefont {Braun}},\ }\bibfield
   {title} {\bibinfo {title} {Perspectives of measuring gravitational effects
  of laser light and particle beams},\ }\href
  {https://doi.org/10.1088/1367-2630/ac5372} {\bibfield  {journal} {\bibinfo
  {journal} {New Journal of Physics}\ }\textbf {\bibinfo {volume} {24}},\
  \bibinfo {pages} {053021} (\bibinfo {year} {2022})}\BibitemShut {NoStop}%
\bibitem [{\citenamefont {Atonga}\ \emph {et~al.}(2024)\citenamefont {Atonga},
  \citenamefont {Martineau}, \citenamefont {Aboushelbaya}, \citenamefont
  {Barrau}, \citenamefont {von~der Leyen}, \citenamefont {Howard},
  \citenamefont {James}, \citenamefont {Lee}, \citenamefont {Lin},
  \citenamefont {Martin}, \citenamefont {Ouatu}, \citenamefont {Paddock},
  \citenamefont {Ruskov}, \citenamefont {Timmis},\ and\ \citenamefont
  {Norreys}}]{PhysRevD.110.044023}%
  \BibitemOpen
  \bibfield  {author} {\bibinfo {author} {\bibfnamefont {E.}~\bibnamefont
  {Atonga}}, \bibinfo {author} {\bibfnamefont {K.}~\bibnamefont {Martineau}},
  \bibinfo {author} {\bibfnamefont {R.}~\bibnamefont {Aboushelbaya}}, \bibinfo
  {author} {\bibfnamefont {A.}~\bibnamefont {Barrau}}, \bibinfo {author}
  {\bibfnamefont {M.}~\bibnamefont {von~der Leyen}}, \bibinfo {author}
  {\bibfnamefont {S.}~\bibnamefont {Howard}}, \bibinfo {author} {\bibfnamefont
  {A.}~\bibnamefont {James}}, \bibinfo {author} {\bibfnamefont
  {J.}~\bibnamefont {Lee}}, \bibinfo {author} {\bibfnamefont {C.}~\bibnamefont
  {Lin}}, \bibinfo {author} {\bibfnamefont {H.}~\bibnamefont {Martin}},
  \bibinfo {author} {\bibfnamefont {I.}~\bibnamefont {Ouatu}}, \bibinfo
  {author} {\bibfnamefont {R.}~\bibnamefont {Paddock}}, \bibinfo {author}
  {\bibfnamefont {R.}~\bibnamefont {Ruskov}}, \bibinfo {author} {\bibfnamefont
  {R.}~\bibnamefont {Timmis}},\ and\ \bibinfo {author} {\bibfnamefont
  {P.}~\bibnamefont {Norreys}},\ }\bibfield  {title} {\bibinfo {title}
  {Gravitational waves from high-power twisted light},\ }\href
  {https://doi.org/10.1103/PhysRevD.110.044023} {\bibfield  {journal} {\bibinfo
   {journal} {Phys. Rev. D}\ }\textbf {\bibinfo {volume} {110}},\ \bibinfo
  {pages} {044023} (\bibinfo {year} {2024})}\BibitemShut {NoStop}%
\bibitem [{\citenamefont {Ribeyre}\ and\ \citenamefont
  {Tikhonchuk}(2012)}]{ribeyre2012high}%
  \BibitemOpen
  \bibfield  {author} {\bibinfo {author} {\bibfnamefont {X.}~\bibnamefont
  {Ribeyre}}\ and\ \bibinfo {author} {\bibfnamefont {V.}~\bibnamefont
  {Tikhonchuk}},\ }\bibfield  {title} {\bibinfo {title} {High frequency
  gravitational waves generation in laser plasma interaction},\ }in\ \href
  {https://doi.org/10.1142/9789814374552_0292} {\emph {\bibinfo {booktitle}
  {The Twelfth Marcel Grossmann Meeting: On Recent Developments in Theoretical
  and Experimental General Relativity, Astrophysics and Relativistic Field
  Theories (In 3 Volumes)}}}\ (\bibinfo {organization} {World Scientific},\
  \bibinfo {year} {2012})\ pp.\ \bibinfo {pages} {1640--1642}\BibitemShut
  {NoStop}%
\bibitem [{\citenamefont {Gelfer}\ \emph {et~al.}(2016)\citenamefont {Gelfer},
  \citenamefont {Kadlecová}, \citenamefont {Klimo}, \citenamefont {Weber},\
  and\ \citenamefont {Korn}}]{10.1063/1.4962520}%
  \BibitemOpen
  \bibfield  {author} {\bibinfo {author} {\bibfnamefont {E.~G.}\ \bibnamefont
  {Gelfer}}, \bibinfo {author} {\bibfnamefont {H.}~\bibnamefont {Kadlecová}},
  \bibinfo {author} {\bibfnamefont {O.}~\bibnamefont {Klimo}}, \bibinfo
  {author} {\bibfnamefont {S.}~\bibnamefont {Weber}},\ and\ \bibinfo {author}
  {\bibfnamefont {G.}~\bibnamefont {Korn}},\ }\bibfield  {title} {\bibinfo
  {title} {Gravitational waves generated by laser accelerated relativistic
  ions},\ }\href {https://doi.org/10.1063/1.4962520} {\bibfield  {journal}
  {\bibinfo  {journal} {Physics of Plasmas}\ }\textbf {\bibinfo {volume}
  {23}},\ \bibinfo {pages} {093107} (\bibinfo {year} {2016})}\BibitemShut
  {NoStop}%
\bibitem [{\citenamefont {Kadlecov{\'a}}\ \emph {et~al.}(2017)\citenamefont
  {Kadlecov{\'a}}, \citenamefont {Klimo}, \citenamefont {Weber},\ and\
  \citenamefont {Korn}}]{kadlecova2017gravitational}%
  \BibitemOpen
  \bibfield  {author} {\bibinfo {author} {\bibfnamefont {H.}~\bibnamefont
  {Kadlecov{\'a}}}, \bibinfo {author} {\bibfnamefont {O.}~\bibnamefont
  {Klimo}}, \bibinfo {author} {\bibfnamefont {S.}~\bibnamefont {Weber}},\ and\
  \bibinfo {author} {\bibfnamefont {G.}~\bibnamefont {Korn}},\ }\bibfield
  {title} {\bibinfo {title} {Gravitational wave generation by interaction of
  high power lasers with matter using shock waves},\ }\href
  {https://doi.org/https://doi.org/10.1140/epjd/e2017-70586-y} {\bibfield
  {journal} {\bibinfo  {journal} {The European Physical Journal D}\ }\textbf
  {\bibinfo {volume} {71}},\ \bibinfo {pages} {1} (\bibinfo {year}
  {2017})}\BibitemShut {NoStop}%
\bibitem [{\citenamefont {Ataman}(2018)}]{PhysRevA.97.063811}%
  \BibitemOpen
  \bibfield  {author} {\bibinfo {author} {\bibfnamefont {S.}~\bibnamefont
  {Ataman}},\ }\bibfield  {title} {\bibinfo {title} {Vacuum birefringence
  detection in all-optical scenarios},\ }\href
  {https://doi.org/10.1103/PhysRevA.97.063811} {\bibfield  {journal} {\bibinfo
  {journal} {Phys. Rev. A}\ }\textbf {\bibinfo {volume} {97}},\ \bibinfo
  {pages} {063811} (\bibinfo {year} {2018})}\BibitemShut {NoStop}%
\bibitem [{\citenamefont {Ahmadiniaz}\ \emph {et~al.}(2025)\citenamefont
  {Ahmadiniaz} \emph {et~al.}}]{Ahmadiniaz_2025}%
  \BibitemOpen
  \bibfield  {author} {\bibinfo {author} {\bibfnamefont {N.}~\bibnamefont
  {Ahmadiniaz}} \emph {et~al.},\ }\bibfield  {title} {\bibinfo {title} {Towards
  a vacuum birefringence experiment at the helmholtz international beamline for
  extreme fields (letter of intent of the biref@hibef collaboration)},\ }\href
  {https://doi.org/10.1017/hpl.2024.70} {\bibfield  {journal} {\bibinfo
  {journal} {High Power Laser Science and Engineering}\ }\textbf {\bibinfo
  {volume} {13}},\ \bibinfo {pages} {e7} (\bibinfo {year} {2025})}\BibitemShut
  {NoStop}%
\bibitem [{\citenamefont {Demkowicz-Dobrzański}\ \emph
  {et~al.}(2015)\citenamefont {Demkowicz-Dobrzański}, \citenamefont
  {Jarzyna},\ and\ \citenamefont {Kołodyński}}]{DEMKOWICZDOBRZANSKI2015345}%
  \BibitemOpen
  \bibfield  {author} {\bibinfo {author} {\bibfnamefont {R.}~\bibnamefont
  {Demkowicz-Dobrzański}}, \bibinfo {author} {\bibfnamefont {M.}~\bibnamefont
  {Jarzyna}},\ and\ \bibinfo {author} {\bibfnamefont {J.}~\bibnamefont
  {Kołodyński}},\ }\bibfield  {title} {\bibinfo {title} {Chapter four -
  quantum limits in optical interferometry}\ }(\bibinfo  {publisher}
  {Elsevier},\ \bibinfo {year} {2015})\ pp.\ \bibinfo {pages}
  {345--435}\BibitemShut {NoStop}%
\bibitem [{\citenamefont {Evans}(2022)}]{evans2022partial}%
  \BibitemOpen
  \bibfield  {author} {\bibinfo {author} {\bibfnamefont {L.~C.}\ \bibnamefont
  {Evans}},\ }\href@noop {} {\emph {\bibinfo {title} {Partial differential
  equations}}},\ Vol.~\bibinfo {volume} {19}\ (\bibinfo  {publisher} {American
  Mathematical Society},\ \bibinfo {year} {2022})\BibitemShut {NoStop}%
\bibitem [{\citenamefont {Polyanin}(2001)}]{polyanin2001handbook}%
  \BibitemOpen
  \bibfield  {author} {\bibinfo {author} {\bibfnamefont {A.~D.}\ \bibnamefont
  {Polyanin}},\ }\href@noop {} {\emph {\bibinfo {title} {Handbook of linear
  partial differential equations for engineers and scientists}}}\ (\bibinfo
  {publisher} {Chapman and hall/crc},\ \bibinfo {year} {2001})\BibitemShut
  {NoStop}%
\bibitem [{\citenamefont {Bastianelli}\ and\ \citenamefont
  {Schubert}(2005)}]{Fiorenzo_Bastianelli_2005}%
  \BibitemOpen
  \bibfield  {author} {\bibinfo {author} {\bibfnamefont {F.}~\bibnamefont
  {Bastianelli}}\ and\ \bibinfo {author} {\bibfnamefont {C.}~\bibnamefont
  {Schubert}},\ }\bibfield  {title} {\bibinfo {title} {One loop photon-graviton
  mixing in an electromagnetic field: part 1},\ }\href
  {https://doi.org/10.1088/1126-6708/2005/02/069} {\bibfield  {journal}
  {\bibinfo  {journal} {Journal of High Energy Physics}\ }\textbf {\bibinfo
  {volume} {2005}},\ \bibinfo {pages} {069} (\bibinfo {year}
  {2005})}\BibitemShut {NoStop}%
\bibitem [{\citenamefont {Bastianelli}\ \emph {et~al.}(2008)\citenamefont
  {Bastianelli}, \citenamefont {Nucamendi}, \citenamefont {Schubert},\ and\
  \citenamefont {Villanueva}}]{Bastianelli_2008}%
  \BibitemOpen
  \bibfield  {author} {\bibinfo {author} {\bibfnamefont {F.}~\bibnamefont
  {Bastianelli}}, \bibinfo {author} {\bibfnamefont {U.}~\bibnamefont
  {Nucamendi}}, \bibinfo {author} {\bibfnamefont {C.}~\bibnamefont
  {Schubert}},\ and\ \bibinfo {author} {\bibfnamefont {V.~M.}\ \bibnamefont
  {Villanueva}},\ }\bibfield  {title} {\bibinfo {title} {Photon–graviton
  mixing in an electromagnetic field},\ }\href
  {https://doi.org/10.1088/1751-8113/41/16/164048} {\bibfield  {journal}
  {\bibinfo  {journal} {Journal of Physics A: Mathematical and Theoretical}\
  }\textbf {\bibinfo {volume} {41}},\ \bibinfo {pages} {164048} (\bibinfo
  {year} {2008})}\BibitemShut {NoStop}%
\bibitem [{\citenamefont {Pezz\'e}\ and\ \citenamefont
  {Smerzi}(2008)}]{PhysRevLett.100.073601}%
  \BibitemOpen
  \bibfield  {author} {\bibinfo {author} {\bibfnamefont {L.}~\bibnamefont
  {Pezz\'e}}\ and\ \bibinfo {author} {\bibfnamefont {A.}~\bibnamefont
  {Smerzi}},\ }\bibfield  {title} {\bibinfo {title} {Mach-zehnder
  interferometry at the heisenberg limit with coherent and squeezed-vacuum
  light},\ }\href {https://doi.org/10.1103/PhysRevLett.100.073601} {\bibfield
  {journal} {\bibinfo  {journal} {Phys. Rev. Lett.}\ }\textbf {\bibinfo
  {volume} {100}},\ \bibinfo {pages} {073601} (\bibinfo {year}
  {2008})}\BibitemShut {NoStop}%
\bibitem [{\citenamefont {Buikema}\ \emph {et~al.}(2020)\citenamefont {Buikema}
  \emph {et~al.}}]{PhysRevD.102.062003}%
  \BibitemOpen
  \bibfield  {author} {\bibinfo {author} {\bibfnamefont {A.}~\bibnamefont
  {Buikema}} \emph {et~al.},\ }\bibfield  {title} {\bibinfo {title}
  {Sensitivity and performance of the advanced ligo detectors in the third
  observing run},\ }\href {https://doi.org/10.1103/PhysRevD.102.062003}
  {\bibfield  {journal} {\bibinfo  {journal} {Phys. Rev. D}\ }\textbf {\bibinfo
  {volume} {102}},\ \bibinfo {pages} {062003} (\bibinfo {year}
  {2020})}\BibitemShut {NoStop}%
\bibitem [{\citenamefont {Marklund}\ and\ \citenamefont
  {Shukla}(2006)}]{RevModPhys.78.591}%
  \BibitemOpen
  \bibfield  {author} {\bibinfo {author} {\bibfnamefont {M.}~\bibnamefont
  {Marklund}}\ and\ \bibinfo {author} {\bibfnamefont {P.~K.}\ \bibnamefont
  {Shukla}},\ }\bibfield  {title} {\bibinfo {title} {Nonlinear collective
  effects in photon-photon and photon-plasma interactions},\ }\href
  {https://doi.org/10.1103/RevModPhys.78.591} {\bibfield  {journal} {\bibinfo
  {journal} {Rev. Mod. Phys.}\ }\textbf {\bibinfo {volume} {78}},\ \bibinfo
  {pages} {591} (\bibinfo {year} {2006})}\BibitemShut {NoStop}%
\bibitem [{\citenamefont {Fedotov}\ \emph {et~al.}(2023)\citenamefont
  {Fedotov}, \citenamefont {Ilderton}, \citenamefont {Karbstein}, \citenamefont
  {King}, \citenamefont {Seipt}, \citenamefont {Taya},\ and\ \citenamefont
  {Torgrimsson}}]{FEDOTOV20231}%
  \BibitemOpen
  \bibfield  {author} {\bibinfo {author} {\bibfnamefont {A.}~\bibnamefont
  {Fedotov}}, \bibinfo {author} {\bibfnamefont {A.}~\bibnamefont {Ilderton}},
  \bibinfo {author} {\bibfnamefont {F.}~\bibnamefont {Karbstein}}, \bibinfo
  {author} {\bibfnamefont {B.}~\bibnamefont {King}}, \bibinfo {author}
  {\bibfnamefont {D.}~\bibnamefont {Seipt}}, \bibinfo {author} {\bibfnamefont
  {H.}~\bibnamefont {Taya}},\ and\ \bibinfo {author} {\bibfnamefont
  {G.}~\bibnamefont {Torgrimsson}},\ }\bibfield  {title} {\bibinfo {title}
  {Advances in qed with intense background fields},\ }\href
  {https://doi.org/https://doi.org/10.1016/j.physrep.2023.01.003} {\bibfield
  {journal} {\bibinfo  {journal} {Physics Reports}\ }\textbf {\bibinfo {volume}
  {1010}},\ \bibinfo {pages} {1} (\bibinfo {year} {2023})},\ \bibinfo {note}
  {advances in QED with intense background fields}\BibitemShut {NoStop}%
\bibitem [{\citenamefont {Karbstein}(2020)}]{particles3010005}%
  \BibitemOpen
  \bibfield  {author} {\bibinfo {author} {\bibfnamefont {F.}~\bibnamefont
  {Karbstein}},\ }\bibfield  {title} {\bibinfo {title} {Probing vacuum
  polarization effects with high-intensity lasers},\ }\href
  {https://doi.org/10.3390/particles3010005} {\bibfield  {journal} {\bibinfo
  {journal} {Particles}\ }\textbf {\bibinfo {volume} {3}},\ \bibinfo {pages}
  {39} (\bibinfo {year} {2020})}\BibitemShut {NoStop}%
\bibitem [{\citenamefont {Ahmadiniaz}\ \emph {et~al.}(2020)\citenamefont
  {Ahmadiniaz}, \citenamefont {Cowan}, \citenamefont {Sauerbrey}, \citenamefont
  {Schramm}, \citenamefont {Schlenvoigt},\ and\ \citenamefont
  {Sch\"utzhold}}]{PhysRevD.101.116019}%
  \BibitemOpen
  \bibfield  {author} {\bibinfo {author} {\bibfnamefont {N.}~\bibnamefont
  {Ahmadiniaz}}, \bibinfo {author} {\bibfnamefont {T.~E.}\ \bibnamefont
  {Cowan}}, \bibinfo {author} {\bibfnamefont {R.}~\bibnamefont {Sauerbrey}},
  \bibinfo {author} {\bibfnamefont {U.}~\bibnamefont {Schramm}}, \bibinfo
  {author} {\bibfnamefont {H.-P.}\ \bibnamefont {Schlenvoigt}},\ and\ \bibinfo
  {author} {\bibfnamefont {R.}~\bibnamefont {Sch\"utzhold}},\ }\bibfield
  {title} {\bibinfo {title} {Heisenberg limit for detecting vacuum
  birefringence},\ }\href {https://doi.org/10.1103/PhysRevD.101.116019}
  {\bibfield  {journal} {\bibinfo  {journal} {Phys. Rev. D}\ }\textbf {\bibinfo
  {volume} {101}},\ \bibinfo {pages} {116019} (\bibinfo {year}
  {2020})}\BibitemShut {NoStop}%
\bibitem [{\citenamefont {Battesti}\ and\ \citenamefont
  {Rizzo}(2012)}]{Battesti_2013}%
  \BibitemOpen
  \bibfield  {author} {\bibinfo {author} {\bibfnamefont {R.}~\bibnamefont
  {Battesti}}\ and\ \bibinfo {author} {\bibfnamefont {C.}~\bibnamefont
  {Rizzo}},\ }\bibfield  {title} {\bibinfo {title} {Magnetic and electric
  properties of a quantum vacuum},\ }\href
  {https://doi.org/10.1088/0034-4885/76/1/016401} {\bibfield  {journal}
  {\bibinfo  {journal} {Reports on Progress in Physics}\ }\textbf {\bibinfo
  {volume} {76}},\ \bibinfo {pages} {016401} (\bibinfo {year}
  {2012})}\BibitemShut {NoStop}%
\bibitem [{\citenamefont {Dunne}()}]{doi:10.1142/9789812775344_0014}%
  \BibitemOpen
  \bibfield  {author} {\bibinfo {author} {\bibfnamefont {G.~V.}\ \bibnamefont
  {Dunne}},\ }\bibinfo {title} {Heisenberg–euler effective lagrangians:
  Basics and extensions},\ in\ \href
  {https://doi.org/10.1142/9789812775344_0014} {\emph {\bibinfo {booktitle}
  {From Fields to Strings: Circumnavigating Theoretical Physics}}},\ pp.\
  \bibinfo {pages} {445--522}\BibitemShut {NoStop}%
\bibitem [{\citenamefont {Dunne}(2012)}]{doi:10.1142/S2010194512007222}%
  \BibitemOpen
  \bibfield  {author} {\bibinfo {author} {\bibfnamefont {G.~V.}\ \bibnamefont
  {Dunne}},\ }\bibfield  {title} {\bibinfo {title} {The heisenberg-euler
  effective action: 75 years on},\ }\href
  {https://doi.org/10.1142/S2010194512007222} {\bibfield  {journal} {\bibinfo
  {journal} {International Journal of Modern Physics: Conference Series}\
  }\textbf {\bibinfo {volume} {14}},\ \bibinfo {pages} {42} (\bibinfo {year}
  {2012})}\BibitemShut {NoStop}%
\bibitem [{\citenamefont {Rikken}\ and\ \citenamefont
  {Rizzo}(2000)}]{PhysRevA.63.012107}%
  \BibitemOpen
  \bibfield  {author} {\bibinfo {author} {\bibfnamefont {G.~L. J.~A.}\
  \bibnamefont {Rikken}}\ and\ \bibinfo {author} {\bibfnamefont
  {C.}~\bibnamefont {Rizzo}},\ }\bibfield  {title} {\bibinfo {title}
  {Magnetoelectric birefringences of the quantum vacuum},\ }\href
  {https://doi.org/10.1103/PhysRevA.63.012107} {\bibfield  {journal} {\bibinfo
  {journal} {Phys. Rev. A}\ }\textbf {\bibinfo {volume} {63}},\ \bibinfo
  {pages} {012107} (\bibinfo {year} {2000})}\BibitemShut {NoStop}%
\bibitem [{\citenamefont {Abu-Shawareb}\ \emph {et~al.}(2024)\citenamefont
  {Abu-Shawareb} \emph {et~al.}}]{PhysRevLett.132.065102}%
  \BibitemOpen
  \bibfield  {author} {\bibinfo {author} {\bibfnamefont {H.}~\bibnamefont
  {Abu-Shawareb}} \emph {et~al.} (\bibinfo {collaboration} {The Indirect Drive
  ICF Collaboration}),\ }\bibfield  {title} {\bibinfo {title} {Achievement of
  target gain larger than unity in an inertial fusion experiment},\ }\href
  {https://doi.org/10.1103/PhysRevLett.132.065102} {\bibfield  {journal}
  {\bibinfo  {journal} {Phys. Rev. Lett.}\ }\textbf {\bibinfo {volume} {132}},\
  \bibinfo {pages} {065102} (\bibinfo {year} {2024})}\BibitemShut {NoStop}%
\bibitem [{\citenamefont {Marklund}\ \emph {et~al.}(2023)\citenamefont
  {Marklund}, \citenamefont {Blackburn}, \citenamefont {Gonoskov},
  \citenamefont {Magnusson}, \citenamefont {Bulanov},\ and\ \citenamefont
  {Ilderton}}]{Marklund_2023}%
  \BibitemOpen
  \bibfield  {author} {\bibinfo {author} {\bibfnamefont {M.}~\bibnamefont
  {Marklund}}, \bibinfo {author} {\bibfnamefont {T.~G.}\ \bibnamefont
  {Blackburn}}, \bibinfo {author} {\bibfnamefont {A.}~\bibnamefont {Gonoskov}},
  \bibinfo {author} {\bibfnamefont {J.}~\bibnamefont {Magnusson}}, \bibinfo
  {author} {\bibfnamefont {S.~S.}\ \bibnamefont {Bulanov}},\ and\ \bibinfo
  {author} {\bibfnamefont {A.}~\bibnamefont {Ilderton}},\ }\bibfield  {title}
  {\bibinfo {title} {Towards critical and supercritical electromagnetic
  fields},\ }\href {https://doi.org/10.1017/hpl.2022.46} {\bibfield  {journal}
  {\bibinfo  {journal} {High Power Laser Science and Engineering}\ }\textbf
  {\bibinfo {volume} {11}},\ \bibinfo {pages} {e19} (\bibinfo {year}
  {2023})}\BibitemShut {NoStop}%
\bibitem [{\citenamefont {Gonoskov}\ \emph {et~al.}(2012)\citenamefont
  {Gonoskov}, \citenamefont {Aiello}, \citenamefont {Heugel},\ and\
  \citenamefont {Leuchs}}]{PhysRevA.86.053836}%
  \BibitemOpen
  \bibfield  {author} {\bibinfo {author} {\bibfnamefont {I.}~\bibnamefont
  {Gonoskov}}, \bibinfo {author} {\bibfnamefont {A.}~\bibnamefont {Aiello}},
  \bibinfo {author} {\bibfnamefont {S.}~\bibnamefont {Heugel}},\ and\ \bibinfo
  {author} {\bibfnamefont {G.}~\bibnamefont {Leuchs}},\ }\bibfield  {title}
  {\bibinfo {title} {Dipole pulse theory: Maximizing the field amplitude from
  $4\ensuremath{\pi}$ focused laser pulses},\ }\href
  {https://doi.org/10.1103/PhysRevA.86.053836} {\bibfield  {journal} {\bibinfo
  {journal} {Phys. Rev. A}\ }\textbf {\bibinfo {volume} {86}},\ \bibinfo
  {pages} {053836} (\bibinfo {year} {2012})}\BibitemShut {NoStop}%
\bibitem [{\citenamefont {Gonoskov}\ \emph {et~al.}(2013)\citenamefont
  {Gonoskov}, \citenamefont {Gonoskov}, \citenamefont {Harvey}, \citenamefont
  {Ilderton}, \citenamefont {Kim}, \citenamefont {Marklund}, \citenamefont
  {Mourou},\ and\ \citenamefont {Sergeev}}]{PhysRevLett.111.060404}%
  \BibitemOpen
  \bibfield  {author} {\bibinfo {author} {\bibfnamefont {A.}~\bibnamefont
  {Gonoskov}}, \bibinfo {author} {\bibfnamefont {I.}~\bibnamefont {Gonoskov}},
  \bibinfo {author} {\bibfnamefont {C.}~\bibnamefont {Harvey}}, \bibinfo
  {author} {\bibfnamefont {A.}~\bibnamefont {Ilderton}}, \bibinfo {author}
  {\bibfnamefont {A.}~\bibnamefont {Kim}}, \bibinfo {author} {\bibfnamefont
  {M.}~\bibnamefont {Marklund}}, \bibinfo {author} {\bibfnamefont
  {G.}~\bibnamefont {Mourou}},\ and\ \bibinfo {author} {\bibfnamefont
  {A.}~\bibnamefont {Sergeev}},\ }\bibfield  {title} {\bibinfo {title} {Probing
  nonperturbative qed with optimally focused laser pulses},\ }\href
  {https://doi.org/10.1103/PhysRevLett.111.060404} {\bibfield  {journal}
  {\bibinfo  {journal} {Phys. Rev. Lett.}\ }\textbf {\bibinfo {volume} {111}},\
  \bibinfo {pages} {060404} (\bibinfo {year} {2013})}\BibitemShut {NoStop}%
\bibitem [{\citenamefont {Efimenko}\ \emph {et~al.}(2019)\citenamefont
  {Efimenko}, \citenamefont {Bashinov}, \citenamefont {Gonoskov}, \citenamefont
  {Bastrakov}, \citenamefont {Muraviev}, \citenamefont {Meyerov}, \citenamefont
  {Kim},\ and\ \citenamefont {Sergeev}}]{PhysRevE.99.031201}%
  \BibitemOpen
  \bibfield  {author} {\bibinfo {author} {\bibfnamefont {E.~S.}\ \bibnamefont
  {Efimenko}}, \bibinfo {author} {\bibfnamefont {A.~V.}\ \bibnamefont
  {Bashinov}}, \bibinfo {author} {\bibfnamefont {A.~A.}\ \bibnamefont
  {Gonoskov}}, \bibinfo {author} {\bibfnamefont {S.~I.}\ \bibnamefont
  {Bastrakov}}, \bibinfo {author} {\bibfnamefont {A.~A.}\ \bibnamefont
  {Muraviev}}, \bibinfo {author} {\bibfnamefont {I.~B.}\ \bibnamefont
  {Meyerov}}, \bibinfo {author} {\bibfnamefont {A.~V.}\ \bibnamefont {Kim}},\
  and\ \bibinfo {author} {\bibfnamefont {A.~M.}\ \bibnamefont {Sergeev}},\
  }\bibfield  {title} {\bibinfo {title} {Laser-driven plasma pinching in
  ${e}^{\ensuremath{-}}{e}^{+}$ cascade},\ }\href
  {https://doi.org/10.1103/PhysRevE.99.031201} {\bibfield  {journal} {\bibinfo
  {journal} {Phys. Rev. E}\ }\textbf {\bibinfo {volume} {99}},\ \bibinfo
  {pages} {031201} (\bibinfo {year} {2019})}\BibitemShut {NoStop}%
\bibitem [{\citenamefont {Bashinov}\ \emph {et~al.}(2022)\citenamefont
  {Bashinov}, \citenamefont {Efimenko}, \citenamefont {Muraviev}, \citenamefont
  {Volokitin}, \citenamefont {Meyerov}, \citenamefont {Leuchs}, \citenamefont
  {Sergeev},\ and\ \citenamefont {Kim}}]{PhysRevE.105.065202}%
  \BibitemOpen
  \bibfield  {author} {\bibinfo {author} {\bibfnamefont {A.~V.}\ \bibnamefont
  {Bashinov}}, \bibinfo {author} {\bibfnamefont {E.~S.}\ \bibnamefont
  {Efimenko}}, \bibinfo {author} {\bibfnamefont {A.~A.}\ \bibnamefont
  {Muraviev}}, \bibinfo {author} {\bibfnamefont {V.~D.}\ \bibnamefont
  {Volokitin}}, \bibinfo {author} {\bibfnamefont {I.~B.}\ \bibnamefont
  {Meyerov}}, \bibinfo {author} {\bibfnamefont {G.}~\bibnamefont {Leuchs}},
  \bibinfo {author} {\bibfnamefont {A.~M.}\ \bibnamefont {Sergeev}},\ and\
  \bibinfo {author} {\bibfnamefont {A.~V.}\ \bibnamefont {Kim}},\ }\bibfield
  {title} {\bibinfo {title} {Particle trajectories, gamma-ray emission, and
  anomalous radiative trapping effects in magnetic dipole wave},\ }\href
  {https://doi.org/10.1103/PhysRevE.105.065202} {\bibfield  {journal} {\bibinfo
   {journal} {Phys. Rev. E}\ }\textbf {\bibinfo {volume} {105}},\ \bibinfo
  {pages} {065202} (\bibinfo {year} {2022})}\BibitemShut {NoStop}%
\bibitem [{\citenamefont {Johnson}(2005)}]{cubature}%
  \BibitemOpen
  \bibfield  {author} {\bibinfo {author} {\bibfnamefont {S.~G.}\ \bibnamefont
  {Johnson}},\ }\href@noop {} {\bibinfo {title} {Multi-dimensional adaptive
  integration in {C}: The {Cubature} package}},\ \bibinfo {howpublished}
  {\url{https://github.com/stevengj/cubature}} (\bibinfo {year}
  {2005})\BibitemShut {NoStop}%
\bibitem [{\citenamefont {Ciufolini}\ \emph {et~al.}(2003)\citenamefont
  {Ciufolini}, \citenamefont {Kopeikin}, \citenamefont {Mashhoon},\ and\
  \citenamefont {Ricci}}]{CIUFOLINI2003101}%
  \BibitemOpen
  \bibfield  {author} {\bibinfo {author} {\bibfnamefont {I.}~\bibnamefont
  {Ciufolini}}, \bibinfo {author} {\bibfnamefont {S.}~\bibnamefont {Kopeikin}},
  \bibinfo {author} {\bibfnamefont {B.}~\bibnamefont {Mashhoon}},\ and\
  \bibinfo {author} {\bibfnamefont {F.}~\bibnamefont {Ricci}},\ }\bibfield
  {title} {\bibinfo {title} {On the gravitomagnetic time delay},\ }\href
  {https://doi.org/https://doi.org/10.1016/S0375-9601(02)01804-2} {\bibfield
  {journal} {\bibinfo  {journal} {Physics Letters A}\ }\textbf {\bibinfo
  {volume} {308}},\ \bibinfo {pages} {101} (\bibinfo {year}
  {2003})}\BibitemShut {NoStop}%
\bibitem [{\citenamefont {Ciufolini}\ and\ \citenamefont
  {Ricci}(2002)}]{Ciufolini_2002}%
  \BibitemOpen
  \bibfield  {author} {\bibinfo {author} {\bibfnamefont {I.}~\bibnamefont
  {Ciufolini}}\ and\ \bibinfo {author} {\bibfnamefont {F.}~\bibnamefont
  {Ricci}},\ }\bibfield  {title} {\bibinfo {title} {Time delay due to spin and
  gravitational lensing},\ }\href {https://doi.org/10.1088/0264-9381/19/15/301}
  {\bibfield  {journal} {\bibinfo  {journal} {Classical and Quantum Gravity}\
  }\textbf {\bibinfo {volume} {19}},\ \bibinfo {pages} {3863} (\bibinfo {year}
  {2002})}\BibitemShut {NoStop}%
\bibitem [{\citenamefont {Mu}\ \emph {et~al.}(2016)\citenamefont {Mu},
  \citenamefont {Li}, \citenamefont {Jing}, \citenamefont {Zhu}, \citenamefont
  {Zhou}, \citenamefont {Wang}, \citenamefont {Zhou}, \citenamefont {Xie},
  \citenamefont {Su}, \citenamefont {Zhang}, \citenamefont {Zeng},
  \citenamefont {Zuo}, \citenamefont {Cao},\ and\ \citenamefont
  {Wang}}]{Mu:16}%
  \BibitemOpen
  \bibfield  {author} {\bibinfo {author} {\bibfnamefont {J.}~\bibnamefont
  {Mu}}, \bibinfo {author} {\bibfnamefont {Z.}~\bibnamefont {Li}}, \bibinfo
  {author} {\bibfnamefont {F.}~\bibnamefont {Jing}}, \bibinfo {author}
  {\bibfnamefont {Q.}~\bibnamefont {Zhu}}, \bibinfo {author} {\bibfnamefont
  {K.}~\bibnamefont {Zhou}}, \bibinfo {author} {\bibfnamefont {S.}~\bibnamefont
  {Wang}}, \bibinfo {author} {\bibfnamefont {S.}~\bibnamefont {Zhou}}, \bibinfo
  {author} {\bibfnamefont {N.}~\bibnamefont {Xie}}, \bibinfo {author}
  {\bibfnamefont {J.}~\bibnamefont {Su}}, \bibinfo {author} {\bibfnamefont
  {J.}~\bibnamefont {Zhang}}, \bibinfo {author} {\bibfnamefont
  {X.}~\bibnamefont {Zeng}}, \bibinfo {author} {\bibfnamefont {Y.}~\bibnamefont
  {Zuo}}, \bibinfo {author} {\bibfnamefont {L.}~\bibnamefont {Cao}},\ and\
  \bibinfo {author} {\bibfnamefont {X.}~\bibnamefont {Wang}},\ }\bibfield
  {title} {\bibinfo {title} {Coherent combination of femtosecond pulses via
  non-collinear cross-correlation and far-field distribution},\ }\href
  {https://doi.org/10.1364/OL.41.000234} {\bibfield  {journal} {\bibinfo
  {journal} {Opt. Lett.}\ }\textbf {\bibinfo {volume} {41}},\ \bibinfo {pages}
  {234} (\bibinfo {year} {2016})}\BibitemShut {NoStop}%
\bibitem [{\citenamefont {Fedotov}\ \emph {et~al.}(2010)\citenamefont
  {Fedotov}, \citenamefont {Narozhny}, \citenamefont {Mourou},\ and\
  \citenamefont {Korn}}]{PhysRevLett.105.080402}%
  \BibitemOpen
  \bibfield  {author} {\bibinfo {author} {\bibfnamefont {A.~M.}\ \bibnamefont
  {Fedotov}}, \bibinfo {author} {\bibfnamefont {N.~B.}\ \bibnamefont
  {Narozhny}}, \bibinfo {author} {\bibfnamefont {G.}~\bibnamefont {Mourou}},\
  and\ \bibinfo {author} {\bibfnamefont {G.}~\bibnamefont {Korn}},\ }\bibfield
  {title} {\bibinfo {title} {Limitations on the attainable intensity of high
  power lasers},\ }\href {https://doi.org/10.1103/PhysRevLett.105.080402}
  {\bibfield  {journal} {\bibinfo  {journal} {Phys. Rev. Lett.}\ }\textbf
  {\bibinfo {volume} {105}},\ \bibinfo {pages} {080402} (\bibinfo {year}
  {2010})}\BibitemShut {NoStop}%
\bibitem [{\citenamefont {Seipt}\ \emph {et~al.}(2017)\citenamefont {Seipt},
  \citenamefont {Heinzl}, \citenamefont {Marklund},\ and\ \citenamefont
  {Bulanov}}]{PhysRevLett.118.154803}%
  \BibitemOpen
  \bibfield  {author} {\bibinfo {author} {\bibfnamefont {D.}~\bibnamefont
  {Seipt}}, \bibinfo {author} {\bibfnamefont {T.}~\bibnamefont {Heinzl}},
  \bibinfo {author} {\bibfnamefont {M.}~\bibnamefont {Marklund}},\ and\
  \bibinfo {author} {\bibfnamefont {S.~S.}\ \bibnamefont {Bulanov}},\
  }\bibfield  {title} {\bibinfo {title} {Depletion of intense fields},\ }\href
  {https://doi.org/10.1103/PhysRevLett.118.154803} {\bibfield  {journal}
  {\bibinfo  {journal} {Phys. Rev. Lett.}\ }\textbf {\bibinfo {volume} {118}},\
  \bibinfo {pages} {154803} (\bibinfo {year} {2017})}\BibitemShut {NoStop}%
\bibitem [{\citenamefont {Radier}\ \emph {et~al.}(2022)\citenamefont {Radier},
  \citenamefont {Chalus}, \citenamefont {Charbonneau}, \citenamefont
  {Thambirajah}, \citenamefont {Deschamps}, \citenamefont {David},
  \citenamefont {Barbe}, \citenamefont {Etter}, \citenamefont {Matras},
  \citenamefont {Ricaud},\ and\ \citenamefont {et~al.}}]{eli_np__2022}%
  \BibitemOpen
  \bibfield  {author} {\bibinfo {author} {\bibfnamefont {C.}~\bibnamefont
  {Radier}}, \bibinfo {author} {\bibfnamefont {O.}~\bibnamefont {Chalus}},
  \bibinfo {author} {\bibfnamefont {M.}~\bibnamefont {Charbonneau}}, \bibinfo
  {author} {\bibfnamefont {S.}~\bibnamefont {Thambirajah}}, \bibinfo {author}
  {\bibfnamefont {G.}~\bibnamefont {Deschamps}}, \bibinfo {author}
  {\bibfnamefont {S.}~\bibnamefont {David}}, \bibinfo {author} {\bibfnamefont
  {J.}~\bibnamefont {Barbe}}, \bibinfo {author} {\bibfnamefont
  {E.}~\bibnamefont {Etter}}, \bibinfo {author} {\bibfnamefont
  {G.}~\bibnamefont {Matras}}, \bibinfo {author} {\bibfnamefont
  {S.}~\bibnamefont {Ricaud}},\ and\ \bibinfo {author} {\bibnamefont
  {et~al.}},\ }\bibfield  {title} {\bibinfo {title} {10 pw peak power
  femtosecond laser pulses at eli-np},\ }\href
  {https://doi.org/10.1017/hpl.2022.11} {\bibfield  {journal} {\bibinfo
  {journal} {High Power Laser Science and Engineering}\ }\textbf {\bibinfo
  {volume} {10}},\ \bibinfo {pages} {e21} (\bibinfo {year} {2022})}\BibitemShut
  {NoStop}%
\end{thebibliography}%
\end{document}